\pdfoutput=1     
\documentclass[12pt]{JINST}
\usepackage{epsfig}
\usepackage{amsbsy}
\usepackage{rotating}
\usepackage{cite}
 \usepackage{textcomp} 
\DeclareFontFamily{U}{euc}{}
\DeclareFontShape{U}{euc}{m}{n}{<-6>eurm5<6-8>eurm7<8->eurm10}{}%
\DeclareSymbolFont{AMSc}{U}{euc}{m}{n} 
\DeclareMathSymbol{\umu}{\mathord}{AMSc}{"16}

\RequirePackage{lineno}
\usepackage{afterpage}
\usepackage{url}

\title{Tracking within Hadronic Showers in the CALICE SDHCAL prototype using a Hough Transform Technique }

\author{\centering 
\LARGE\bf The CALICE Collaboration
}

\author{\centering
Z.\,Deng,
Y.\,Li,
Y.\,Wang, 
Q.\,Yue, 
Z.\,Yang
\\ \it
Tsinghua University, Department of Engineering Physics, Beijing, 100084, P.R.
China
}
\author{\centering
D.\,Boumediene, C.\,Carloganu, V.\,Fran\c cais
\\ \it
Universit\'e Clermont Auvergne, Universit\'e Blaise Pascal, Universit\'e Blaise Pascal, CNRS/IN2P3, LPC, 4 Av. Blaise Pascal, TSA/CS 60026,
F-63178 Aubi\`ere, France
}
\author{\centering 
G.\,Cho, 
D-W.\,Kim, 
S.\,C.\,Lee, 
Z.\, Liu,
W.\,Park, 
S.\,Vallecorsa
\\ \it
Gangneung-Wonju National University, Gangnung 25457, South Korea
}
\author{\centering
S.\,Cauwenbergh, 
M.\,Tytgat,
A.\,Pingault,
N.\,Zaganidis
\\ \it
Ghent University, Department of Physics and Astronomy,
Proeftuinstraat 86, B-9000 Gent, Belgium
}

 \author{\centering
O.\,Bach,
E.\,Brianne,
A.\,Ebrahimi,
K.\,Gadow,
P.\,G\"{o}ttlicher,
O.\,Hartbrich$^1$,
A.\,Irles,
K.\,Kotera$^2$,
F.\,Krivan,
K.\,Kr\"{u}ger,
S.\,Lu,
C.\,Neub\"{u}ser$^3$,
A.\,Provenza,
M.\,Reinecke,
F.\,Sefkow,
S.\,Schuwalow,
Y.\,Sudo,
H.L.\,Tran
\\ \it
DESY, Notkestrasse 85,
D-22603 Hamburg, Germany
} 
\author{\centering
H.\,Hirai,
K.\,Kawagoe,
T.\,Suehara,
H.\,Sumida,
T.\,Yoshioka,
\\ \it
Department of Physics and Research Center for Advanced Particle Physics,
Kyushu University, 744 Motooka, Nishi-ku, Fukuoka 819-0395, Japan
} 
\author{\centering 
E.\,Cortina Gil,
S.\,Mannai
\\ \it
Center for Cosmology, Particle Physics and Cosmology (CP3)
Universit\'{e} catholique de Louvain, Chemin du cyclotron 2,
1320 Louvain-la-Neuve, Belgium
}
\author{\centering 
V.\,Buridon,
C.\,Combaret, 
L.\,Caponetto,
R.\,Et\'{e}, 
G.\,Garillot,
G.\,Grenier, 
R.\,Han$^4$,
J.C.\,Ianigro,
R.\,Kieffer$^5$,
T. \,Kurca, 
I.\,Laktineh,
B.\,Li,
N.\,Lumb, 
H.\,Mathez, 
L.\,Mirabito, 
A.\,Petrukhin$^6$,
A.\,Steen$^7$
\\ \it
Univ. Lyon, Universit\'{e} Lyon 1, 
CNRS/IN2P3, IPNL 4 rue E Fermi 69622,
Villeurbanne CEDEX, France
}
\author{\centering 
J.\,Berenguer~Antequera,
E.\,Calvo~Alamillo, 
M.-C.\,Fouz, 
J.\,Marin,
J.\,Navarrete,
J.\,Puerta-Pelayo, 
A.\,Verdugo
\\ \it
CIEMAT, Centro de Investigaciones Energeticas, Medioambientales y Tecnologicas, Madrid, Spain 
}
\author{\centering
F.\,Corriveau
\\ \it
Department of Physics, McGill University,
Ernest Rutherford Physics Bldg.,
3600 University Ave.,
Montr\'{e}al, Qu\'{e}bec,
Canada H3A 2T8
} 
\author{\centering 
M.\,Chadeeva$^8$
\\ \it
P.\,N.\, Lebedev Physical Institute,
Russian Academy of Sciences,
117924 GSP-1 Moscow, B-333, Russia
}

\author{\centering
M.\,Gabriel,
P.\,Goecke,
C.\,Graf,
Y.\,Israeli,
N.\,van der Kolk,
F.\,Simon,
M.\,Szalay,
H.\,Windel
\\ \it
Max-Planck-Institut f\"ur Physik, F\"ohringer Ring 6, D-80805 Munich, Germany
}
\author{\centering
S.\,Bilokin,
J.\,Bonis,
R.\,P\"oschl,
A.\,Thiebault,
F.\,Richard,
D.\,Zerwas
\\ \it
Laboratoire de l'Acc\'elerateur Lin\'eaire,
CNRS/IN2P3 et Universit\'e de Paris-Sud XI,
Centre Scientifique d'Orsay B\^atiment 200,
BP 34,
F-91898 Orsay CEDEX, France
} 
 \author{\centering
M.\,Anduze,
V.\,Balagura,
E.\,Becheva,
V.\,Boudry,
J-C.\,Brient,
R.\,Cornat,
F.\,Gastaldi,
Y.\,Haddad$^{9}$,
F.\,Magniette,
J.\,Nanni,
 M.\,Ruan$^{10}$,
M.\,Rubio-Roy,
K.\,Shpak,
T.H.\,Tran,
H.\,Videau,
D.\,Yu${^{11}}$
\\ \it
Laboratoire Leprince-Ringuet (LLR) -- \'{E}cole Polytechnique,
CNRS/IN2P3,
Palaiseau, F-91128 France
} 
\author{\centering   
S.\,Callier,
F.\,Dulucq, 
Ch.\,de la Taille, 
G.\,Martin-Chassard, 
L.\,Raux, 
N.\,Seguin-Moreau
\\  \it
 Laboratoire OMEGA  -- \'{E}cole Polytechnique,
 CNRS/IN2P3,
 Palaiseau, F-91128 France
}
\author{\centering
J.\,Cvach,
M.\,Janata,
M.\,Kovalcuk,
J.\,Kvasnicka$^{12}$,
I.\,Polak,
J.\,Smolik,
V.\,Vrba,
J.\,Zalesak,
J.\,Zuklin
\\ \it
Institute of Physics, The Czech Academy of Sciences,
Na Slovance 2, CZ-18221 Prague 8, Czech Republic
}

\author{\\
\llap{$^1$} {Now at University of Hawaii, Manoa.}\\
\llap{$^2$}{Now at xy University.}\\
\llap{$^3$}{Now at CERN.}\\
\llap{$^4$}{Now at CAST, Beijing.}\\
\llap{$^5$}{Now at CERN.}\\
\llap{$^6$}{Now at Siegen University, Siegen.}\\
\llap{$^7$}{Now at the National Taiwan University, Taipei.}\\
\llap{$^8$}{Also at MEPhI.}\\
\llap{$^{9}$}{Now at Imperial College, London.}\\
 \llap{$^{10}$}{Now at IHEP, Beijing.}\\
 \llap{$^{11}$}{Also at at IHEP, Beijing.}\\
 \llap{$^{12}$}{Also at DESY.}\\
}
\author{
\it
$^\star$ Corresponding authors\newline
E-mail: \email {laktineh@in2p3.fr,a.petrukhin@cern.ch, a.steen@cern.ch}
}

\abstract{ The high granularity of the  CALICE Semi-Digital Hadronic CALorimeter (SDHCAL) provides the capability  to reveal the track segments present in hadronic showers. These segments are then used as a tool to probe the behaviour of the active layers in situ, to better reconstruct the energy of these hadronic showers and also to distinguish them from electromagnetic ones. In addition, the comparison of these track segments in data and  the simulation helps to discriminate among the different shower models used in the simulation.   To extract the track segments in the  showers recorded in the SDHCAL, a Hough Transform is used after  being adapted to the presence of the dense core of the hadronic showers and the SDHCAL active medium structure.}

\begin{document}


\section{Introduction}
Hadronic showers produced by the interaction of hadrons in the absorber part of a sampling calorimeter like the CALICE SDHCAL prototype~\cite{Prototype} often contain several track segments associated to charged particles. Some of these particles cross several active layers before being stopped or reacting inelastically and others could even escape the calorimeter. High-granularity calorimeters provide an excellent tool to reconstruct such track segments. These segments could be used to monitor the active layers of the calorimeters in situ. They allow a better understanding of the response of the calorimeter and consequentially a better  estimation of the hadronic energy. In addition, they help in the comparaison of the different hadronic shower models used in the simulation~\cite{AHCALSegment}.
The capability of the calorimeter to construct such tracks depends on its granularity. The higher the granularity, the lower the confusion between such track segments and the remaining parts of the hadronic shower. Still, even with a highly-granular calorimeter, it is difficult to separate the track segments  from the highly dense environment present in showers produced by high-energy hadrons. To achieve such a separation efficiently in the SDHCAL we propose to use  the Hough Transform (HT) method\cite{HT-paper-1} that was developed  more than half a century ago to find tracks in a noisy environment. \\ 
In this paper we present the results obtained by applying the HT method to extract track segments in hadronic showers collected during the exposure of the SDHCAL prototype to hadron beams at the H2 CERN SPS beam line in 2012, and we show how these track segments can be used to study the response of the active medium of the calorimeter in situ and how they can help to better understand the hadronic shower structure to improve on the estimation of its energy.  The paper is organized as follows: sect.~2 introduces the HT method  and explains how the method is applied in the case of a dense environment such as that encountered in hadronic showers.  
In sect.~3  the use of the HT track segments as a tool to monitor the efficiency and pad multiplicity of the active layers of the hadronic calorimeter is presented.
Their  exploitation to improve on the hadronic energy resolution and to separate electromagnetic from hadronic showers is also discussed. In sect.~4, we show how  the method  is applied to estimate the number of track segments, their length and direction with respect to the incoming hadron in the data collected by the SDHCAL prototype and in simulated events using a few models. A comparison among these models and the data is presented.

\section{ Hough Transform tracking in the SDHCAL}

\subsection{Hough Transform method}

The Hough Transform is a simple and reliable method that allows aligned points to be recognised among scattered points and the parameters of the straight line joining them to be reconstructed. The scope of this method is larger than what is presented here but all the variants of this method are based on the same principle. To search for points located on a straight line in a plane (say $(z,x)$), the Cartesian coordinates of each point present in the plane are used to define a curve in the associated polar plane $(\theta, \rho)$\cite{HT-paper-2}:

\begin{center}
\begin{equation}
\rho = z \cos{\theta} +  x \sin{\theta}
\end{equation}
\label{eq1}
\end{center}

\begin{figure}[!ht]
\includegraphics[width=.5\textwidth]{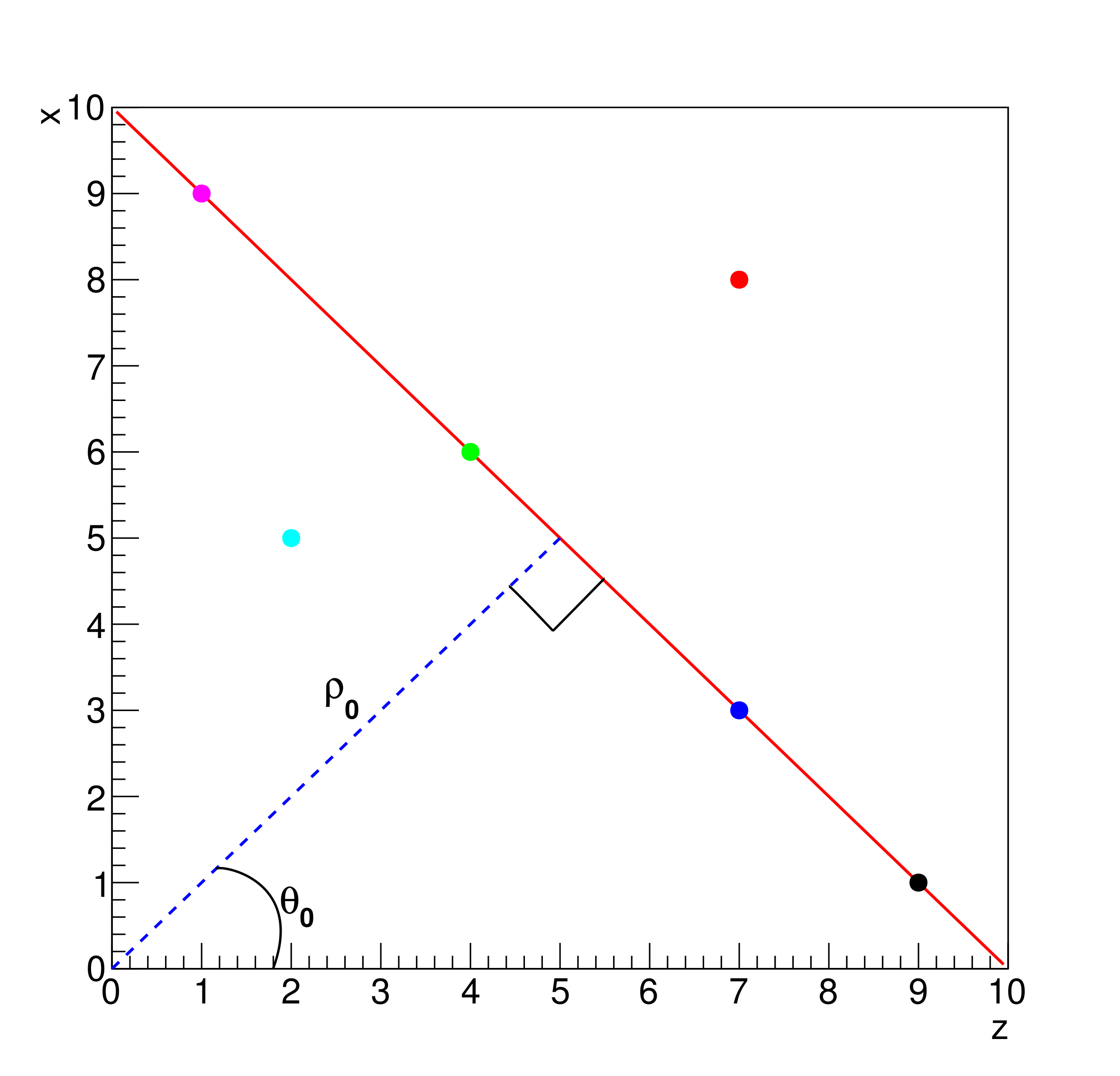}
\hfill
\includegraphics[width=.5\textwidth]{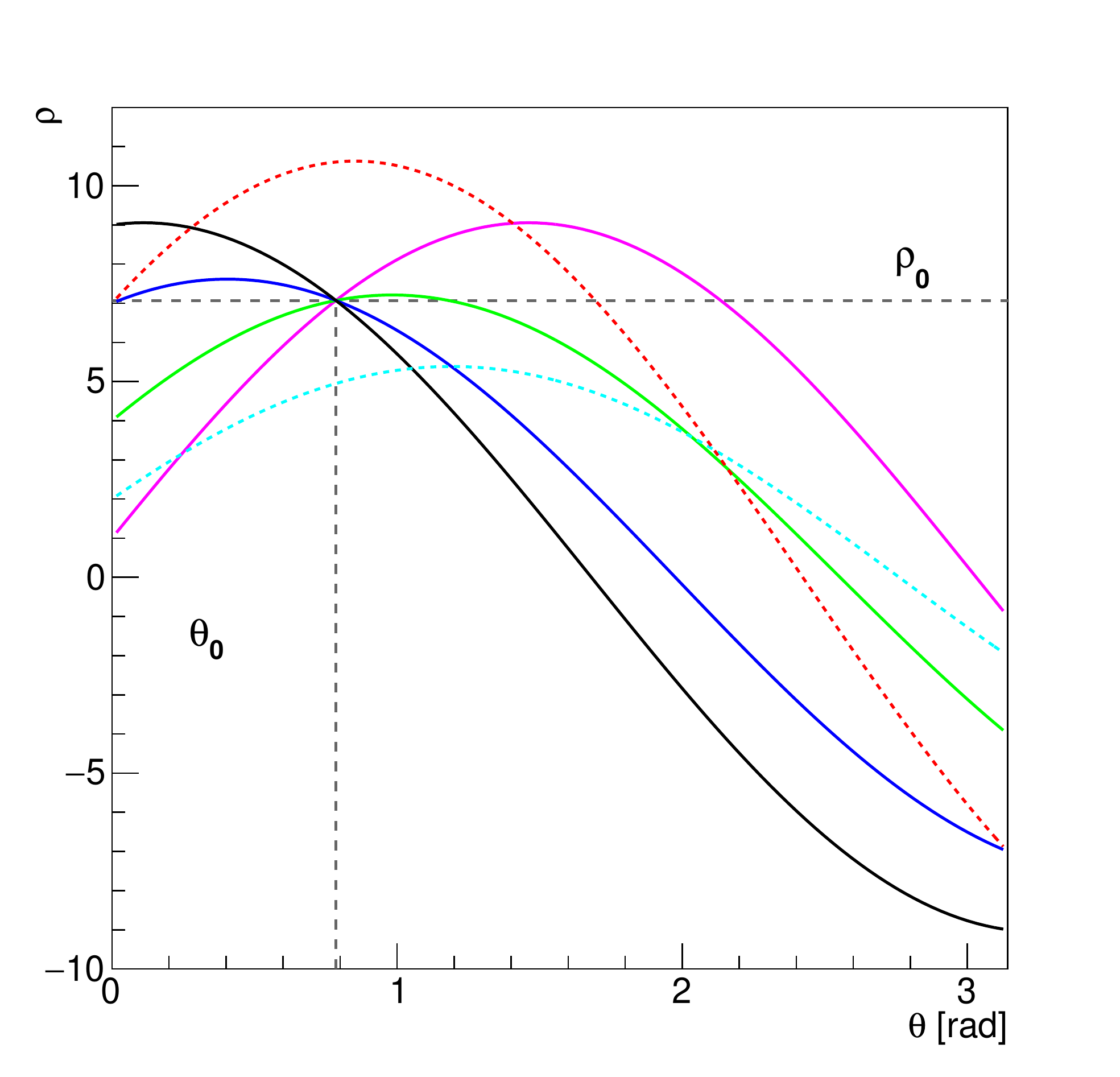}
\caption{Illustration of the Hough Transform method. Each point on the left figure has its associated curve on the right one with the same colour. The curves associated to the points located on the straight line, intersect in the same point of the $(\theta, \rho)$ plane. Arbitrary units are used here for the illustration.}
\label{HT-demo}
\end{figure}

 Under this transformation aligned points have their curves intersecting at one node $(\theta^0, \rho^0)$  of the polar plane. The node's coordinates in the polar plane  determine the angle of the straight line with respect to the ordinate axis of the $(z,x)$ plane and the distance of the straight line to its origin point as can be seen in figure~\ref{HT-demo}.  
Curves associated to the other points intersect also with those of the aligned ones. In scenarios with low density of points outside the track segments, these intersections  scarcely coincide with  $(\theta^0, \rho^0)$.  Therefore, to find aligned points  one should look for nodes in the $(\theta, \rho)$ plane. The number of intersecting curves in one node is an essential parameter to estimate the number of aligned points. In a dense environment the same method can be applied. However, one should take into account here the possibility that the points contributing to one node do not all necessarily belong to a given track.\\

The method described above cannot be applied as such to the hits (fired pads) left by particles in a pixellated detector. The spatial resolution of the hit positions in the detector and more importantly the Multiple Coulomb Scattering (MCS) of the associated charged particles requires the method to be adapted accordingly. This can be achieved by discretising the $(\theta, \rho)$ plane in a 2D histogram. For this histogram, the bins are incremented  each time they are crossed by a curve associated to one point in the $(z,x)$ plane. The size of the bins reflects the expected resolution associated to the hits belonging to one particle and can be optimized to help finding not only straight tracks  but also slightly curved ones. In this way the nodes of intersecting curves are replaced by bins whose content depends on the number of aligned hits.\\

The Hough Transform technique was proposed and successfully used in muon tracking in high energy physics~\cite{{HTT_1}, {HTT_2}}.  Here we propose to use it in calorimetry. 

\subsection { Hough Transform in a hadronic shower}

The SDHCAL is made of 48 active layers interleaved with absorber plates, 2 cm thick each,  made of stainless steel. Each active layer is segmented into $96\times 96$ pickup pads of 1~cm$\times$1~cm. Two adjacent pads are separated by $0.406$ mm. Hits left by charged particles in one layer are represented by fired pads. Each of the pads is read out thanks to an independent electronic channel providing a semi-digital information. Indeed, following the amount of charge induced in a pad, one, two or three  thresholds could be crossed.  The threshold values are fixed by software. They are chosen to indicate roughly if the pad is crossed by one, a few or many charged particles in order to improve on the hadronic shower energy reconstruction as explained in ref.~\cite{sdhcal-paper}. A pad is fired if at least the lowest of the three thresholds is crossed.  A small number of pads are fired  by the passage of a single charged particle in an active layer. The average value of this number is called the  pad multiplicity.\\  

Hadronic showers are generally characterized by a dense core (electromagnetic part) located in the centre and a less dense part (the hadronic one) in the periphery.  To use the Hough Transform to find tracks within hadronic showers one should avoid using hits located in the dense core  since one can build artificially many tracks from the numerous hits of the core.  This can be achieved by keeping only hits that have a small number of neighbours in the same layer.  To apply the Hough Transform to the hadronic showers collected in the SDHCAL prototype, a system of coordinates is needed. In this paper the $z$  axis is chosen parallel to the beam and $x$ and $y$ are the horizontal and vertical axes parallel to the prototype layers.
In order to  reduce computing time and improve efficiency, the Hough Transform method is not applied  to the hits themselves but to the  clusters of hits resulting from aggregating  topologically  neighbouring hits in one active layer $(x,y)$ plane.  These clusters are built recursively.  The first cluster is built starting from the first hit of the collection of hits found in a given layer.  Adjacent hits sharing an edge or an angle with this first hit are looked for and are added to the cluster.  Hits adjacent to any of these hits are again added.  The procedure is applied until no new adjacent hit is found. The hits belonging to this cluster are tagged and withdrawn from the hit list. The same procedure is applied to a new hit chosen among the remaining ones of the plane and repeated until all the hits are gathered into clusters.

The coordinates of  a resulting cluster are that of  the geometrical barycentre of its hits.  Of all the clusters only those with less than 5 hits are kept since clusters belonging to track segments are expected to have at most 4 hits in the $(x,y)$ plane. To eliminate the contribution of the shower's dense part, clusters are rejected if they have more than 2~neighbouring clusters in an area of 10~cm$\times$10~cm around or if  one cluster with more than 5 hits is found in this area.

To look for  3D track segments our method is based on applying the Hough Transform to the $(z,x)$ coordinates of all the remaining clusters to find the  candidate segments in the plane $(z,x)$. A second Hough Transform iteration is then applied to the $(z,y)$ coordinates of the hits associated to each of the candidate segments found in the first step. This allows the elimination of hits accidentally aligned with the track segment's hits in one plane while reconstructing  3D track segments and find their straight line parameters without going through the technically complicated 3D application of the Hough Transform as explained in the following:


The clusters are used to fill a 2D histogram with 100 bins in the range $0 < \theta < \pi $  and 150 bins in the range  $0 < \rho < 150$  cm.  From the $(z,x)$  coordinates of each cluster, the $\rho$  value for each bin of $\theta$ is  computed using eq.~\ref{eq1} and an entry is made in the histogram.

After filling the histogram, bins with more than 6 entries are considered.  
The choice of 6 entries is a compromise  between finding tracks long enough to perform efficiency studies and at the same time keeping tracks associated to low energy particles affected by multiple scattering as well as those produced by particles undergoing inelastic interactions.

Depending on the histogram binning, the neigbouring  bins of the one where the curves intersect may also be crossed by more than six of these curves and are thus also selected. To find the right bin, only the most populated among adjacent bins is kept.   

To eliminate the scenario of accidentally aligned clusters in the 2D $(z,x)$ plane  and not in the 3D $(z,x,y)$ space, the $(z,y)$ coordinates of the clusters belonging to one selected bin are used to fill a second histogram $(\theta^\prime, \rho^\prime)$ with the same binning as the one used for $(\theta, \rho)$ histogram. If any  of the  $(\theta^\prime, \rho^\prime)$ bins is found with more than 6 entries then among all the previous clusters, only those which contribute to this bin are kept.   The clusters found in this way are then used to build a track segment whose parameters are determined from the four values : $ \theta, \rho, \theta^\prime, \rho^\prime$.  It is worth mentioning here that using the $(z,y)$ plane followed by the $(z,x)$ plane gives similar results. 

Finally, to eliminate tracks made of topologically uncorrelated  clusters,  each of the clusters associated to a one selected bin is  compared to the other clusters of the same bin.   The cluster is kept if at least two other clusters from the track are found in the 3 adjacent layers located before and after that of the considered cluster. This requirement is fulfilled in 99\% of the cases for genuine tracks in the SDHCL.


\subsection{Hough-Transform tracks within hadronic showers in the SDHCAL}
\begin{figure}[!ht]
\includegraphics[width=.5\textwidth]{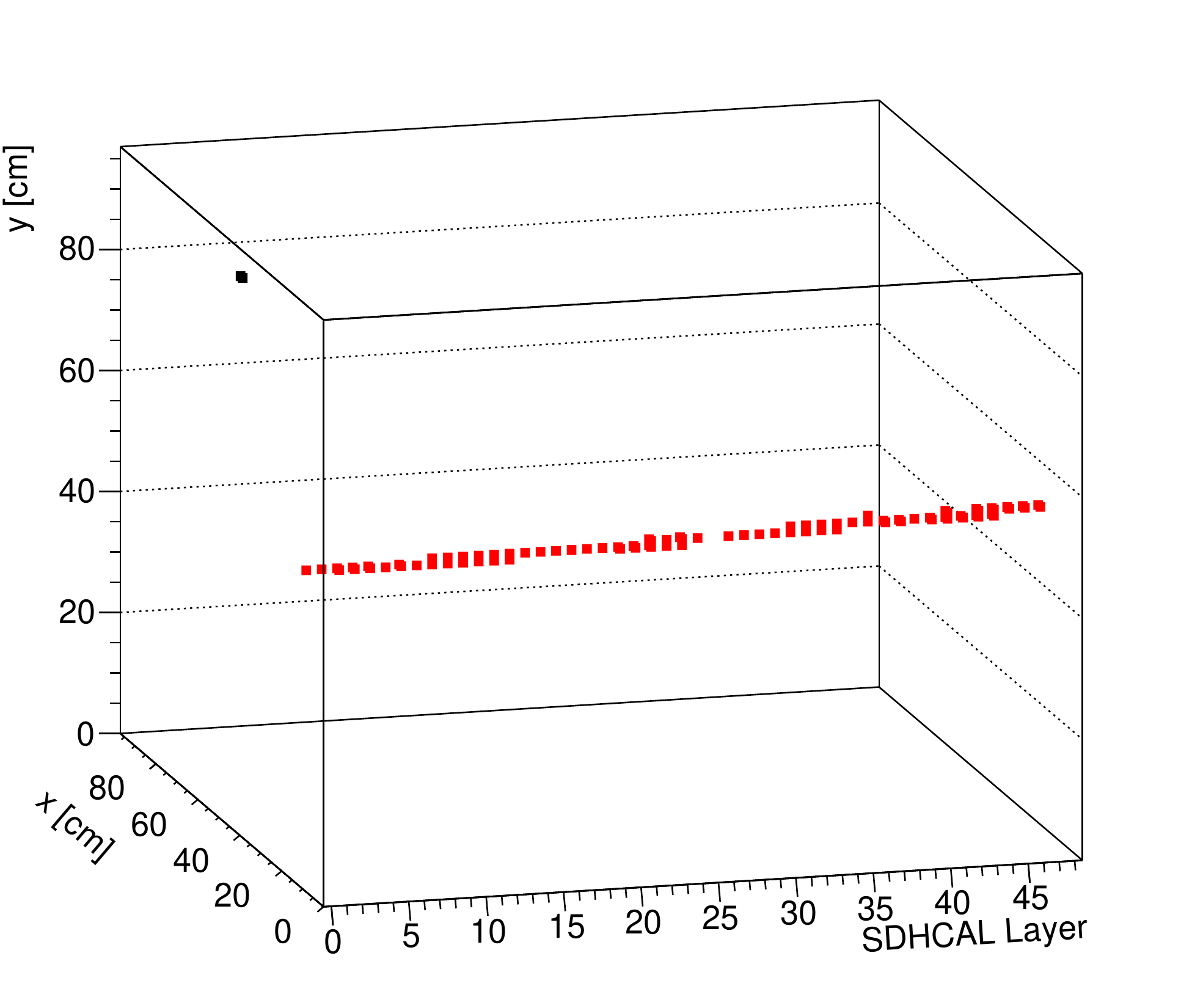}
\hfill
\includegraphics[width=.4\textwidth]{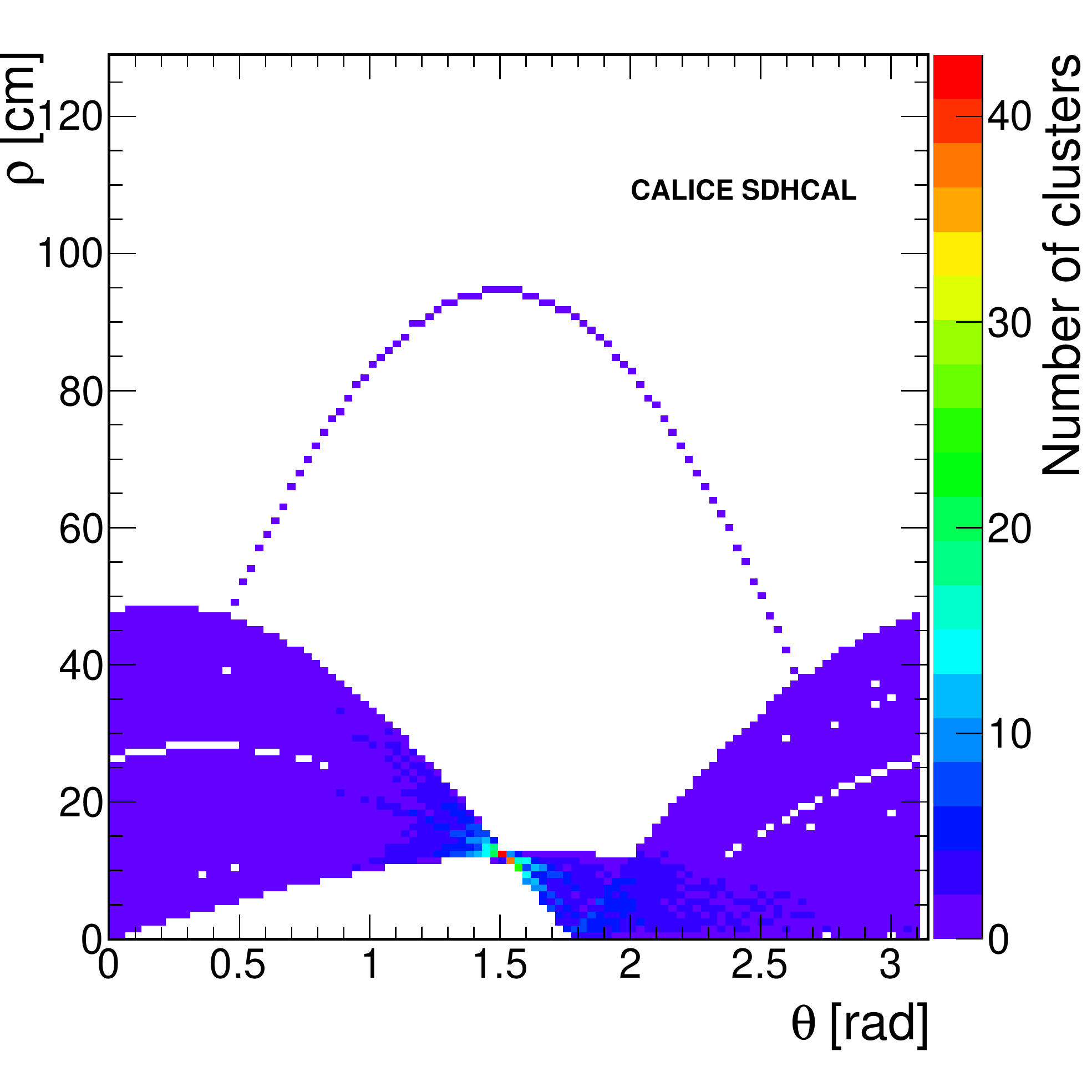}
\caption{Left : Event display of a 50 GeV muon in the SDHCAL. The muon is impinging the SDHCAL perpendicularly to the $(x,y)$ plane at a point around ($x$ =12~cm, $y$ = 53~cm). Red points are the identified hits belonging to a track while the black point  is not. 
Right : Associated $(\theta,\rho)$ histogram which corresponds to the transformation of the $(z,x)$ plane.}
\label{fig:muon}
\end{figure}
The Hough Transform method described in the previous section is first applied  to cosmic and beam-muon events collected in the SDHCAL prototype during its exposure to particle beams at the CERN SPS in 2012.   Most of the hits belonging to these events  are found to be well selected while noisy hits outside the particle path are not as is shown in figure~\ref{fig:muon}.

The Hough Transform method is then applied to events containing showers produced by negative pions.  The auto-trigger  mode used in operating the SDHCAL prototype leads to data containing all kinds of events.  Electrons, cosmic and beam muons contaminating the hadronic events were rejected using  topology-based criteria as described in~\cite{sdhcal-paper}. The noise contribution was also analyzed and found to be of the order of one hit per event which has negligible effect on the present study.
  
 The energies of the collected hadronic showers cover a range going from 10 to 80~GeV. Many of the track segments  of the hadronic showers  are associated with  low energy particles that undergo multiple Coulomb scattering but are still able to cross a few active layers. To select such tracks one can either select Hough Transform histogram bins with a low number of clusters or increase the $\theta$ bin size of this histogram. The last option is technically simpler to  apply and was therefore chosen.  Although the algorithm used here is not optimized to reduce the CPU time, it is found that an average time of $0.16\pm0.08$~s is needed to analyze one hadronic shower of 80~GeV.\\

\begin{figure}[!ht]
\includegraphics[width=.5\textwidth]{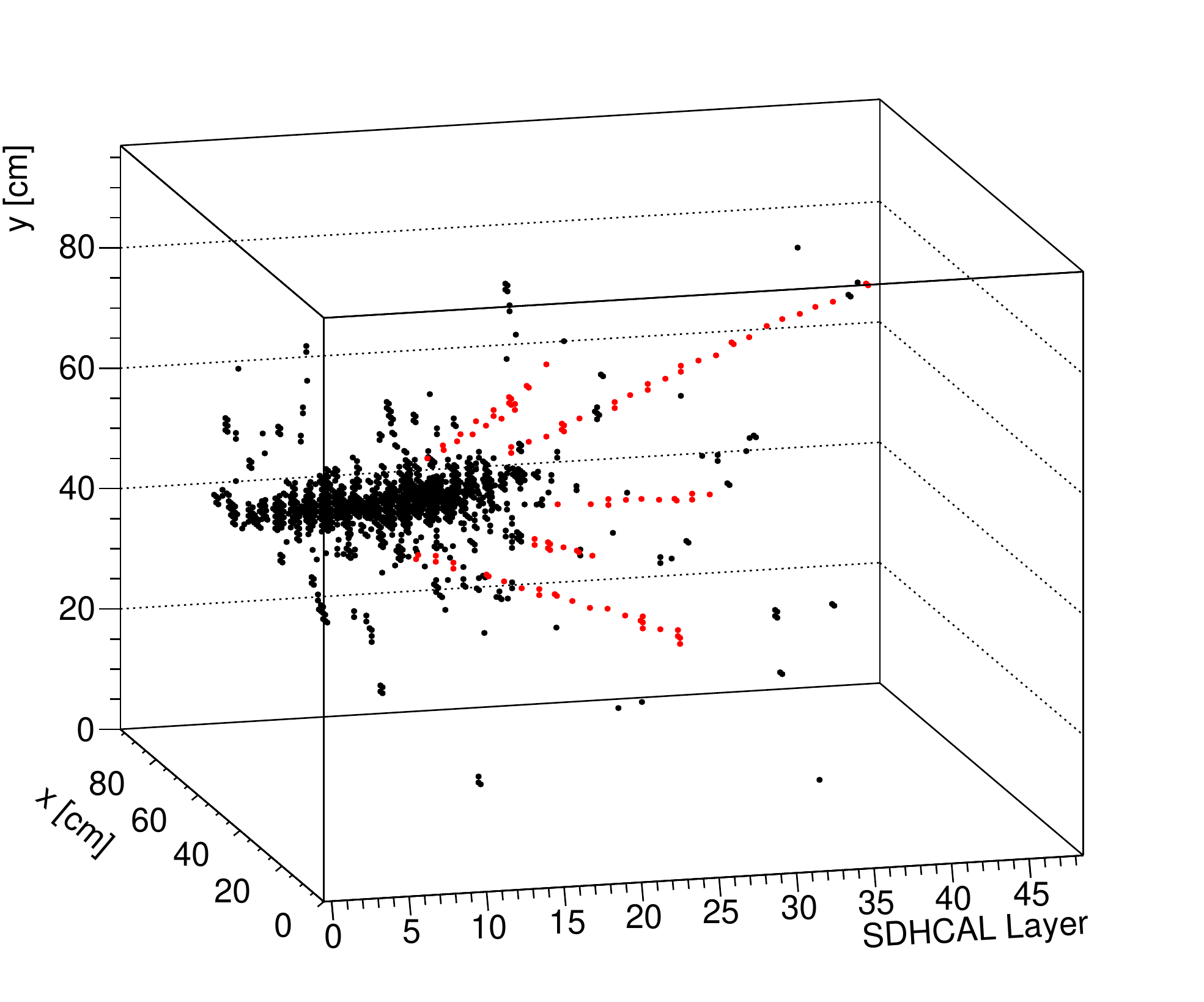}
\hfill
\includegraphics[width=.5\textwidth]{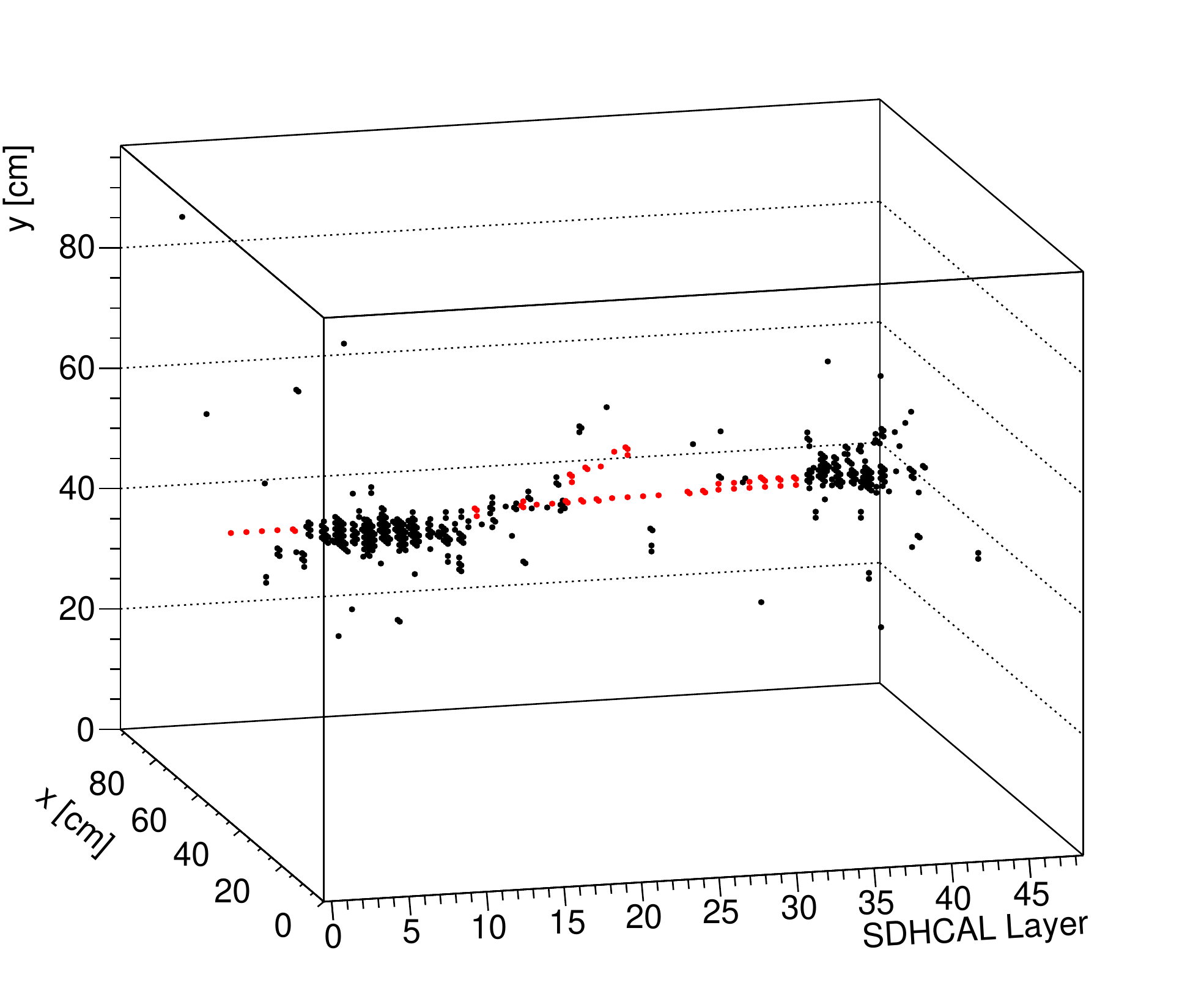}
\caption{Event display of 80 GeV (left) and 30 GeV (right) hadronic showers with hits belonging to tracks in red.}
\label{fig:showers}
\end{figure}

As an illustration of the method  two event displays of hadronic showers at 30 and 80 GeV are shown in figure~\ref{fig:showers} with the Hough Transform selected hits. 

\section{ Use of Hough Transform tracks in the SDHCAL}
\subsection{Calibration purposes and PFA applications}

\begin{figure}[!ht]
\begin{center}
\includegraphics[width=.45\textwidth]{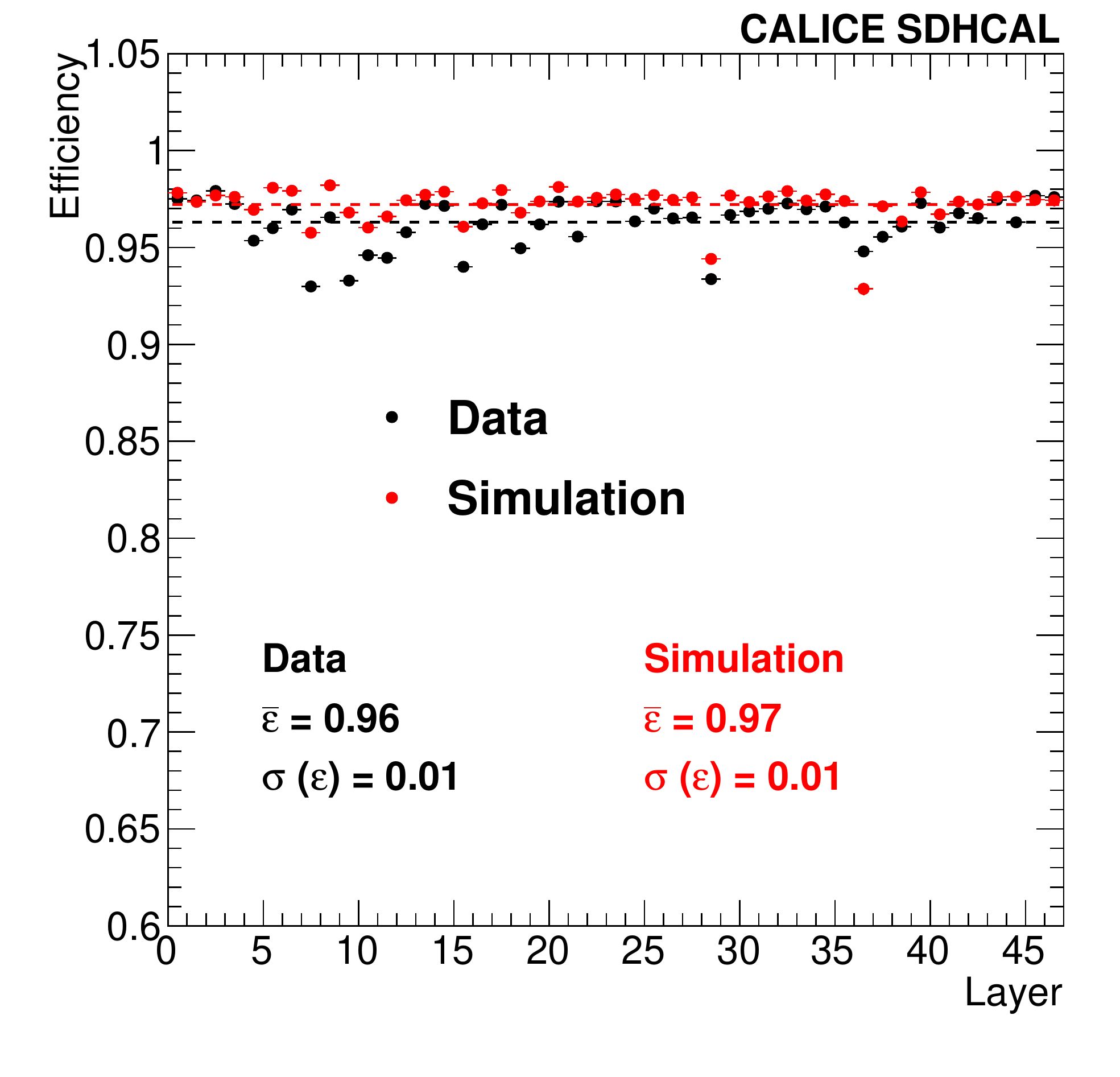}
\includegraphics[width=.45\textwidth]{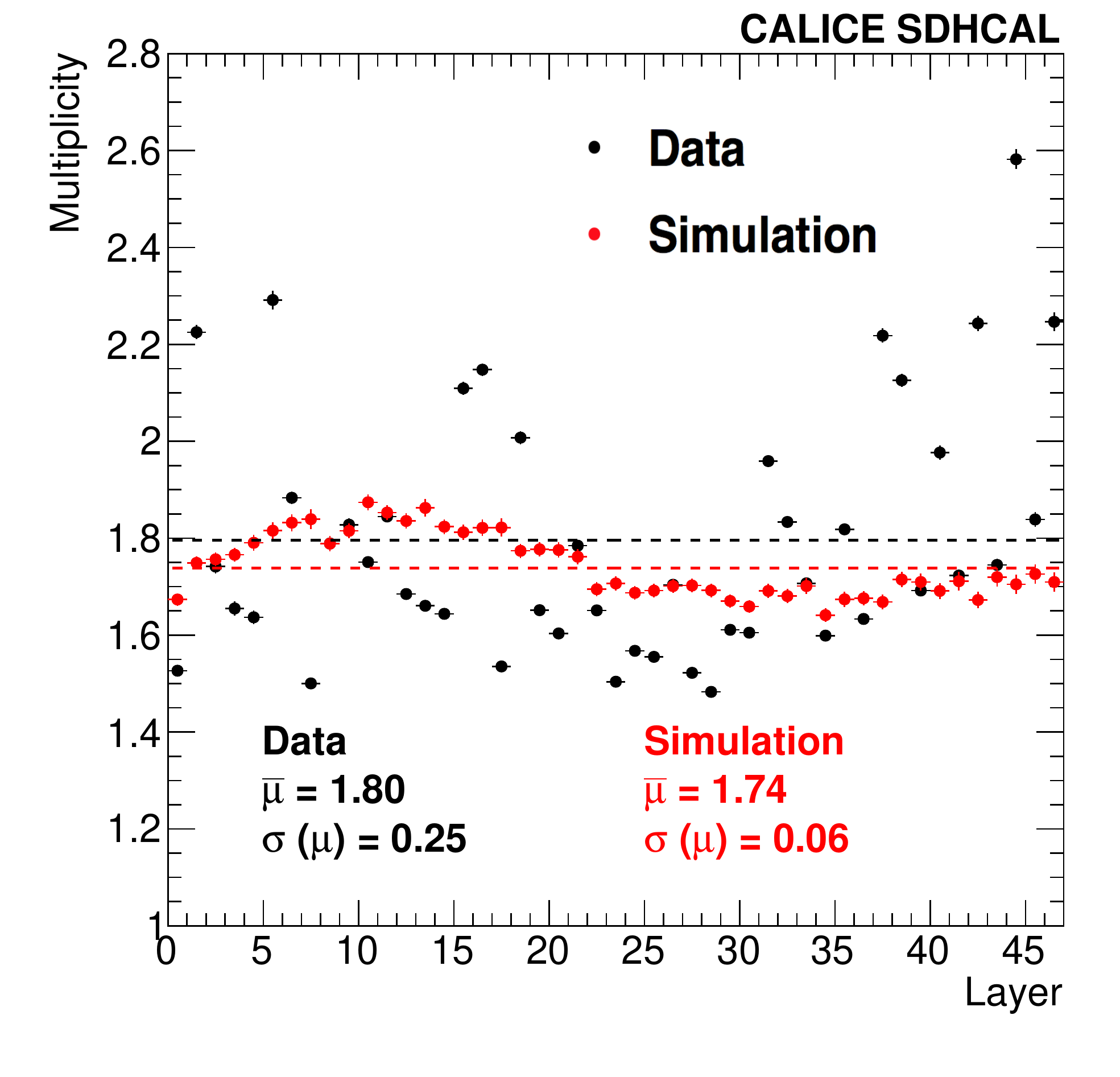}
\caption{Efficiency and multiplicity per layer for a 40 GeV pion in data (black dots) and in the simulation (red dots). Zones instrumented  by  faulty electronic ASICs are excluded. Values of the average efficiency $(\bar \epsilon)$ and of the average hit multiplicity ($\bar \mu$) as well as the  RMS of the efficiency distribution $(\sigma(\epsilon))$ and  that of the hit multiplicity distribution $(\sigma(\mu))$ of the 48 layers are also given.}
\label{fig:eff-multi-per-layer}
\end{center}
\end{figure}

The tracks one can extract from hadronic showers play an important role in checking the active layer behaviour in situ by studying the efficiency and multiplicity of the detector. To achieve such  a study,  high quality Hough Transform segments are selected. Indeed, the aligned clusters are fitted to straight lines in both the $(z,x)$ and $(y,x)$ planes using a least-$\chi^2$ technique. The variance ($\sigma^2$)  used in the $\chi^2$ definition here is the transverse cluster size (number of  pads in the cluster along $x$ or $y$) divided by 12. Only tracks with  $\chi^2/\hbox{NdF} <1$ with $\hbox{NdF = N}_l$ -2 are chosen, where $\hbox{N}_l$ is the number of layers containing the clusters used to fit a straight line of 2 unknown parameters. To study one layer, only clusters belonging to the other layers are kept. 
The intercept of  the straight line associated with the segment in the studied layer is determined. The efficiency of this layer is then estimated by looking for clusters in a 2~cm radius around the intercept. If at least one cluster is found,  the multiplicity is then estimated by counting the number of hits associated to the closest cluster.  Average efficiency and multiplicity per layer using the events collected in a 40 GeV pion run and that of simulated pion events of the same energy are shown in figure~\ref{fig:eff-multi-per-layer}. These results are generally consistent with what was observed in ref.~\cite{sdhcal-paper} where those two variables were estimated with beam muons.  The slight difference, systematically positive, compared with the results obtained with beam muons (at the level of a few percent) is related to the fact that contrary to the latter the angle of the track segments in the hadronic shower is not necessarily perpendicular to the SDHCAL layers. This angular difference results in slightly higher efficiency and pad multiplicity with respect to that  observed in cosmic muons~\cite{digitizer}.

\begin{figure}[!ht]
\begin{center}
\includegraphics[width=.5\textwidth]{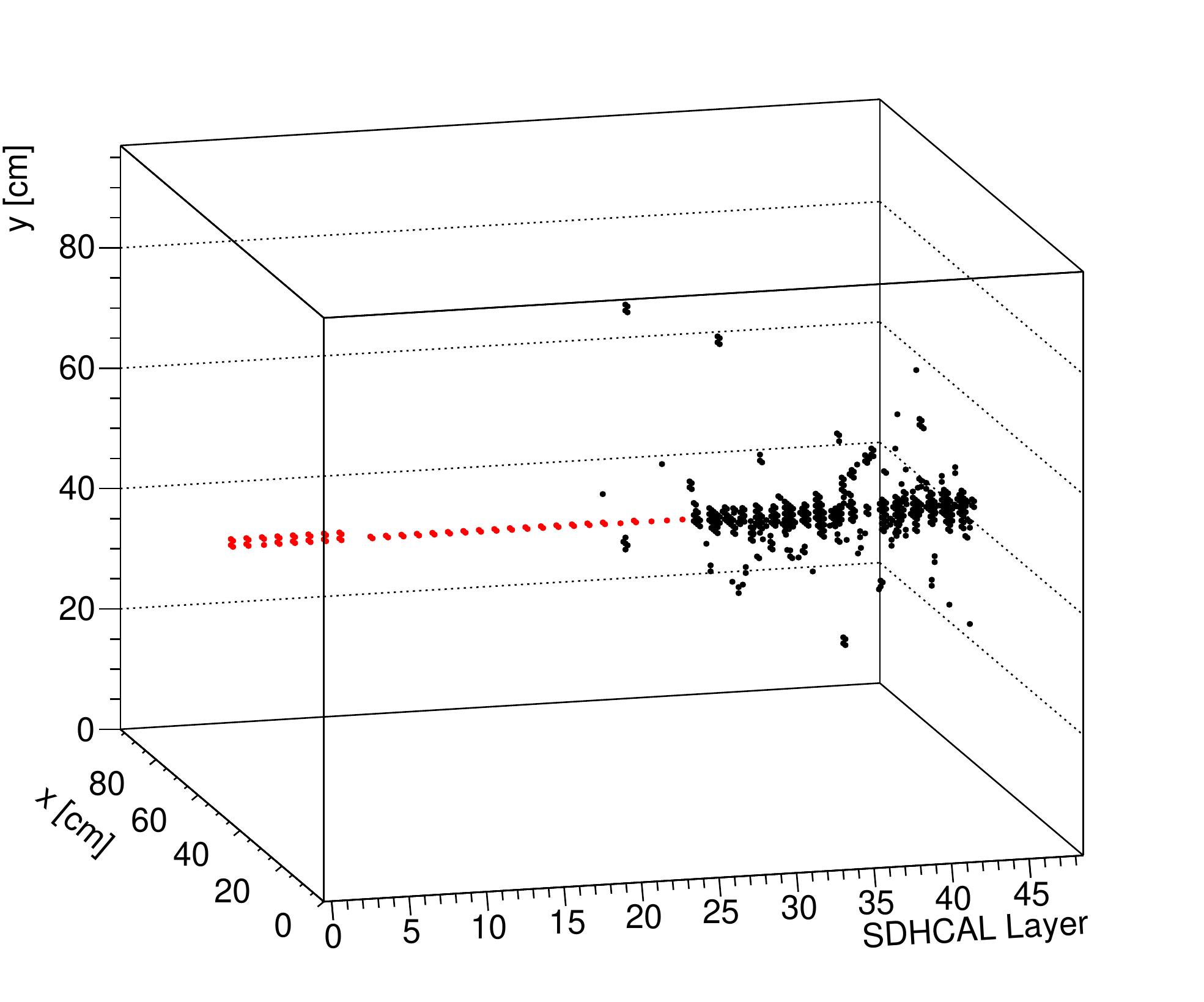}
\caption{Event display of a 80 GeV pion in the SDHCAL. The track (in red) before the shower starts, is well identified and could be useful for PFA.}
\label{fig:PFA}
\end{center}
\end{figure}

\begin{figure}[!ht]
\begin{center}
\includegraphics[width=.50\textwidth]{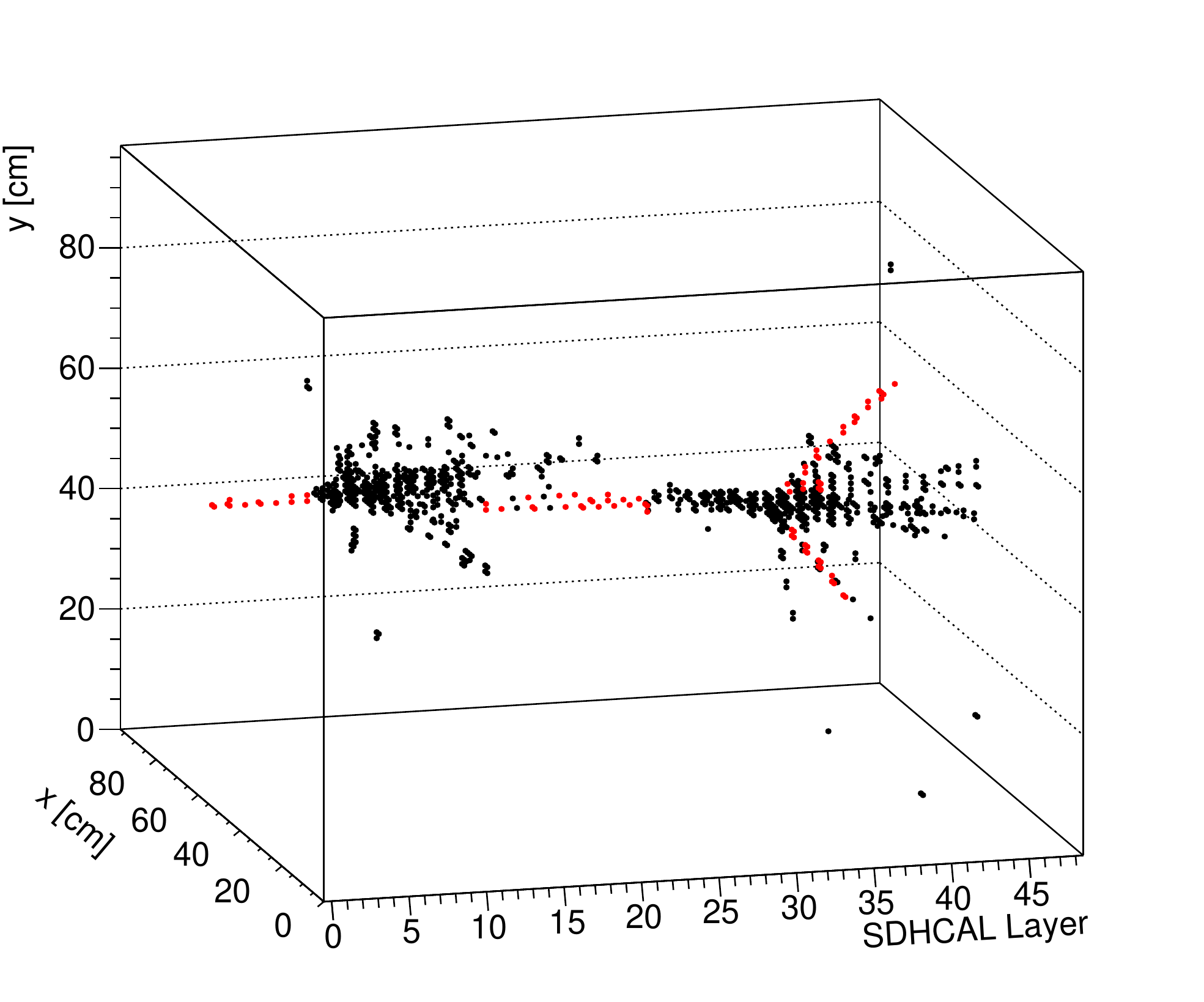}
\caption{50 GeV hadronic shower illustrating that connection between clusters could be done with the reconstructed tracks.}
\label{fig:connection}
\end{center}
\end{figure}

In addition to the use of  track segments in hadronic showers to check the behaviour of the SDHCAL active layers, they could be useful in Particle Flow Algorithm (PFA) techniques~\cite{{PFA_1}, {PFA_2}, {PFA_3}}.  Indeed, a  by-product of this method is the  possibility to tag the track segment associated to the incoming charged particle in the calorimeter as can be seen in figure~\ref{fig:PFA}. This is an interesting feature that could be exploited  in PFA techniques where the contribution of charged hadrons is to be separated from that  of neutral ones. If a tracker is placed in front of the calorimeter,  the connection between the track segment in the calorimeter and that of the tracker leads to a  better estimation of the charged hadron energy  by using its momentum which is often more precisely measured in the tracker.\\
Track segments could  also be used to separate nearby hadronic showers by connecting clusters produced by hadronic interactions of secondary charged particles to the main one as can be seen in figure~\ref{fig:connection}. A successful association increases the probability to attach correctly the clusters to the right  particle  reducing thus  the confusion between  the charged and neutral hadrons and preventing a possible  energy double-counting.   To quantify however  the real contribution of using such track segments in improving the PFA performance a detailed study is needed. \\
\subsection{Hadronic shower energy reconstruction}

 Another potential advantage of the track segments is to improve on the energy reconstruction.  In the SDHCAL energy reconstruction method, each of the thresholds is given a different weight~\cite{sdhcal-paper} to account for the number of tracks crossing one pad. The reconstructed energy is estimated as 
 
 \begin{center}
\begin{equation}
 E^{\mathrm{}}_{\mathrm{reco}} = \alpha N_1 + \beta N_2 + \gamma N_3 
 \label{eq3}
\end{equation}
\end{center}     
 where $N_1$ corresponds to  the number of hits which are above the first threshold and below the second, $N_2$ denotes the number of hits  which are above both the first and the second but below the third threshold and  $N_3$  the number of hits that are above the third threshold. $ \alpha, \beta$ and $\gamma$ are parameterized as quadratic functions of the total number of hits  ($N_\mathrm{hit} =    N_1+N_2+N_3$).

Tracks of low energy that stop inside the calorimeter may have hits passing the second or the third threshold, especially those located at the end of the segment\footnote{The high ionisation value d$E$/d$x$ of the tracks at the end produces more charges and thus hits  of the higher thresholds are often observed.}. These hits of a single track segment may bias the energy estimation based on this method. Therefore, giving the same weight for all the hits belonging to these track segments could improve on the energy reconstruction. To check this assumption, the same procedure of energy reconstruction for hits others than those selected by the HT method is applied and a constant weight  is assigned to the latter as follows:
  
\begin{center}
\begin{equation}
 {E^{\mathrm{HT}}_{\mathrm{reco}} = \alpha' N'_1 + \beta' N'_2 + \gamma' N'_3 + c N_\mathrm{HT}}
\end{equation}
\label{eq2}
\end{center}      

where $N_\mathrm{HT}$ is the number of hits belonging to track segments selected by the HT method. $N'_1, N'_2$ and $N'_3$ are respectively  $N_1, N_2$ and $N_3$  after subtracting the hits belonging to track segments.  $ \alpha', \beta'$ and $\gamma'$ are new quadratic functions of the total number of hits  ($N_\mathrm{hit} =    N'_1+N'_2+N'_3+N_\mathrm{HT}$).  
 
\begin{figure}[!ht]
\begin{center}
\includegraphics[width=.70\textwidth]{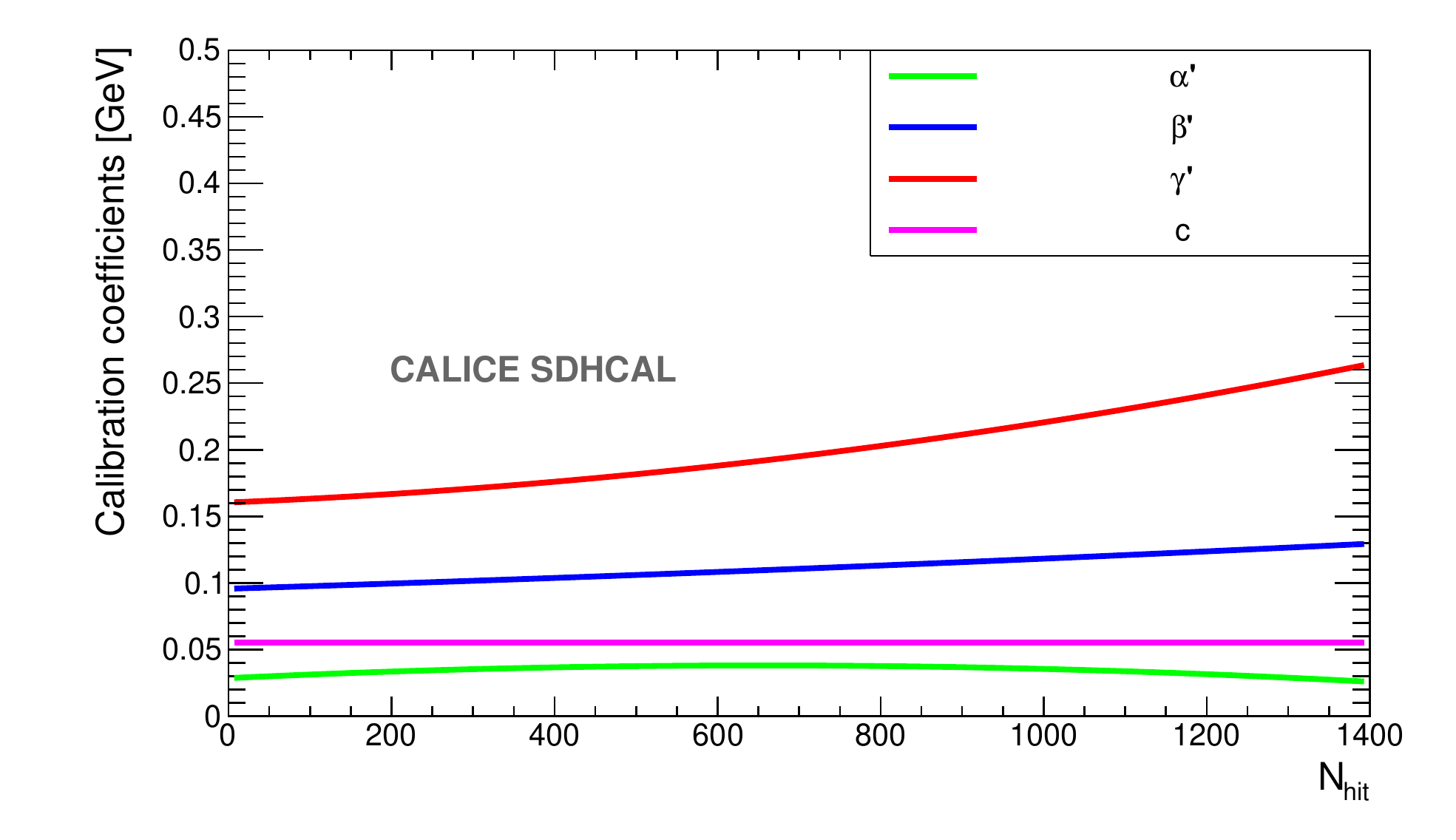}
\caption{Evolution of the coefficients $\alpha'$ (green), $\beta'$ (blue),
$\gamma'$ (red) and $c$ (magenta) in terms of the total number of hits
$N_\mathrm{hit}$. }
\label{fig:evolution}
\end{center}
\end{figure}

A $\chi^2$-like optimization procedure similar to the one described in ref.\cite{sdhcal-paper} is then performed to determine the nine parameters associated to $ \alpha', \beta'$ and $\gamma'$ functions as well as the parameter $c$. The evolution of  $ \alpha', \beta'$ and $\gamma'$ as a function of $N_\mathrm{hit}$ as well as the constant $c$ is presented in figure~\ref{fig:evolution}.  

The energy of the  hadronic events collected in the H2 test beam in 2012 is then estimated using the new formula of eq.~\ref{eq2}. To estimate the energy resolution the same recipe of~\cite{sdhcal-paper} is applied. First, a Gaussian is used to fit over the whole full range of the distribution. Then, a Gaussian is fitted only in the range of $\pm 1.5 \sigma$ of the mean value of the first fit. The $\sigma$ of the second fit is used as the energy resolution  R(E$^{\mathrm{HT}}_{\mathrm{reco}}$) and the new mean value as the reconstructed energy E$^{\mathrm{HT}}_{\mathrm{reco}}$.  The relative energy resolution is thus given by the ratio R(E$^{\mathrm{HT}}_{\mathrm{reco}}$)/ E$^{\mathrm{HT}}_{\mathrm{reco}}$.  

\begin{figure}[!ht]
\begin{center}
\includegraphics[width=.5\textwidth]{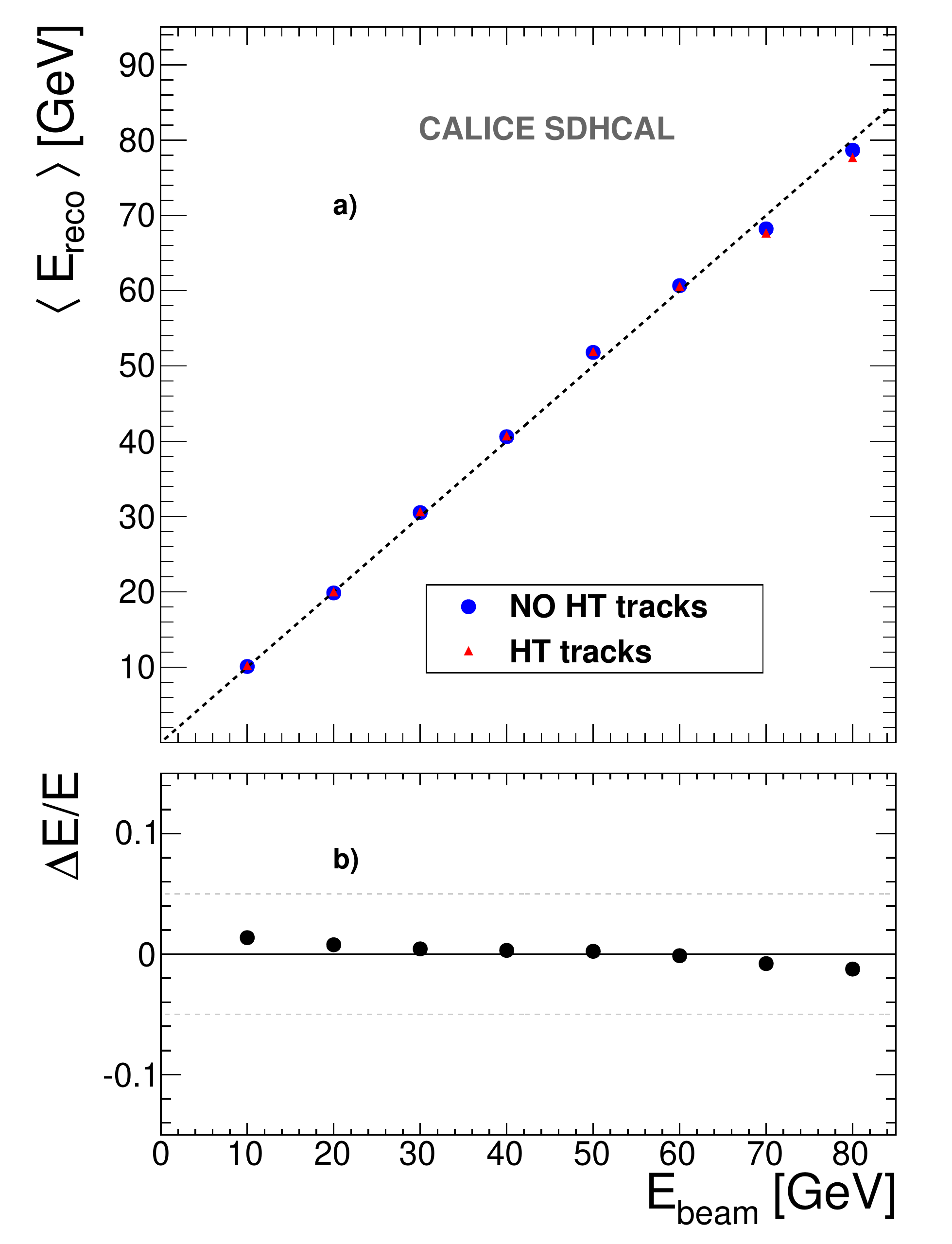}
\caption{ a) Reconstructed energy using the standard method as presented in ref.~\cite{sdhcal-paper} and the one using the HT track segments as a function of the beam energy. b) The relative difference of reconstructed energy $\Delta$E/E= ((E$^{\mathrm{HT}}_{\mathrm{reco}}$-E$^{\mathrm{NoHT}}_{\mathrm{reco}}$)/ E$^{\mathrm{NoHT}}_{\mathrm{reco}}$) for both methods as a function of the beam energy.  
Both statistical and systematic uncertainties are included in the error bars that are  within the marker size.}
\label{fig:linearity-comparison}
\end{center}
\end{figure}

\begin{figure}[!ht]
\begin{center}
\includegraphics[width=.5\textwidth]{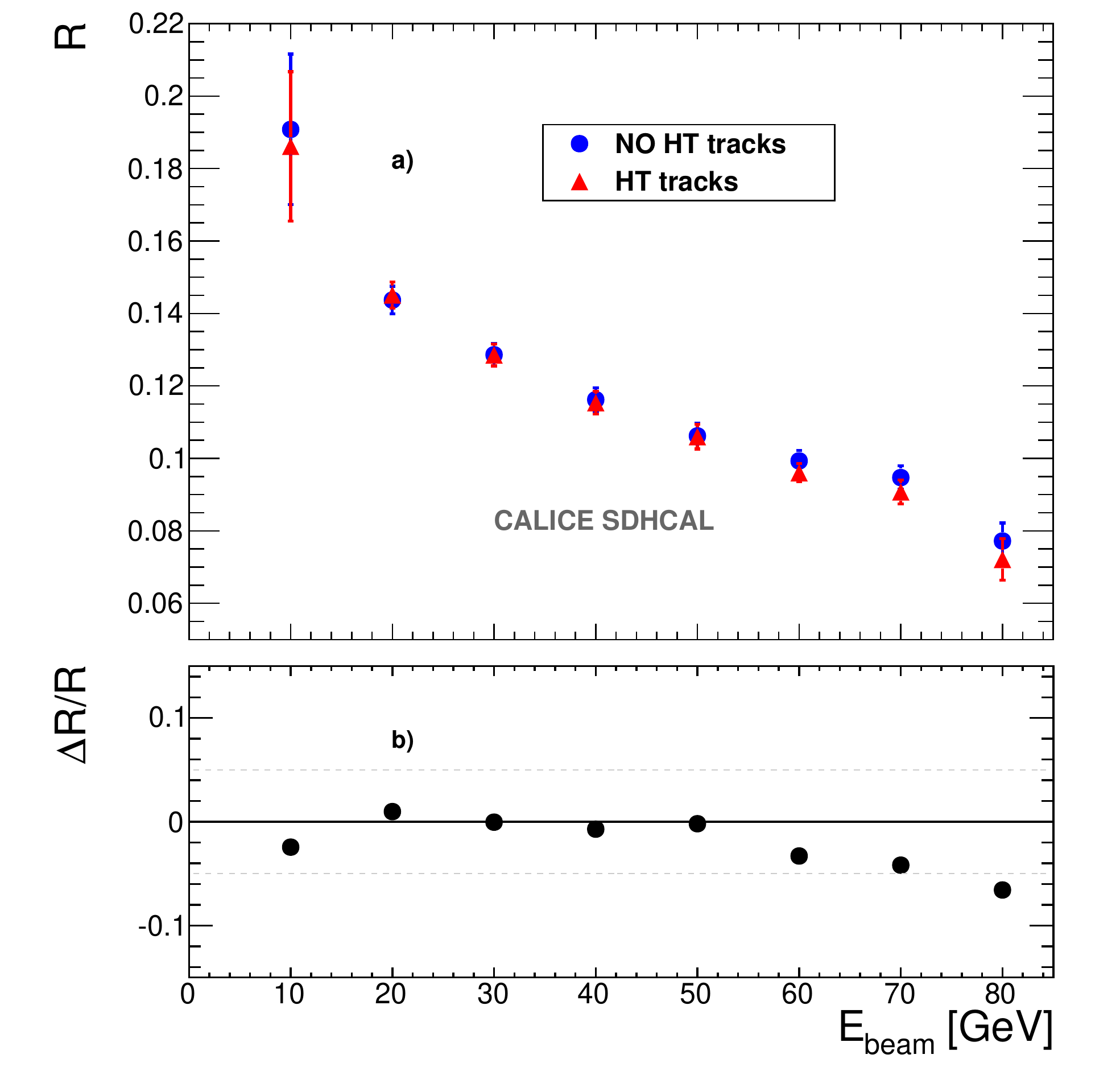}
\caption{a) Resolution R of the reconstructed energy using the standard method as presented in ref.~\cite{sdhcal-paper} and the one using the HT track segments as a function of the beam energy. b) The relative difference of reconstructed energy resolution $\Delta$R/R  = (R(E$^{\mathrm{HT}}_{\mathrm{reco}})$-R( E$^{\mathrm{NoHT}}_{\mathrm{reco}}$) /  R(E$^{\mathrm{NoHT}}_{\mathrm{reco}}$)) as a function of the beam energy.  Both statistical and systematic uncertainties are included in the error bars. They are  within the marker size in the bottom plot.}
\label{fig:resolution-comparison}
\end{center}
\end{figure}

Results are then compared with those obtained in ref.~\cite{sdhcal-paper}. The reconstructed energy obtained using the two methods as a function of the beam energy is shown in figure~\ref{fig:linearity-comparison} (top).  The relative difference of the two is  also shown  in figure~\ref{fig:linearity-comparison} (bottom). Good linearity is obtained with the two methods. Figure~\ref{fig:resolution-comparison} shows the energy resolution with the two methods as well as the relative difference.  
 At energies higher than 40~GeV, where the second and third  thresholds play an important role as explained in ref.~\cite{sdhcal-paper}, assigning the same weight to hits of  track segments independently of their threshold improves the energy resolution by a few percent though it makes the linearity slightly worse.  In addition, the higher the energy the more the number of track segments produced in hadronic showers as will be shown in section~\ref{models} which explains why the improvement is enhanced with the energy.   Finally, the fact that  the second and third threshold hits  of the track segments represent on average about  5 {\textperthousand}  of the total number of hits  and that the contribution to the energy reconstruction of these high-threshold hits  based on eq.~\ref{eq3}  is a few times higher than the one they have by using  eq.~\ref{eq2}, as can be seen in  figure~\ref{fig:evolution},  could  explain the relative improvement of a few percent observed in applying the new method.

Statistical and systematic uncertainties are included in the results shown in the previous figures.  The sources of the systematic uncertainties included in this study are the same as those detailed in ref.~\cite{sdhcal-paper}. These sources are the ones related to the method used to estimate the energy resolution, the hadronic events selection criteria, the effect of the beam intensity and the uncertainty of the beam energy.     

At low energy, the systematic uncertainties related to the measurement of the resolution and to the event selection are of the same order and dominate, while at high energy, the one due to the beam intensity correction represents about half of the total uncertainty.   The uncertainty on the beam energy was found to be negligible in all cases.


\subsection{Electromagnetic and hadronic shower separation}
\begin{figure}[!ht]
\includegraphics[width=.32\textwidth]{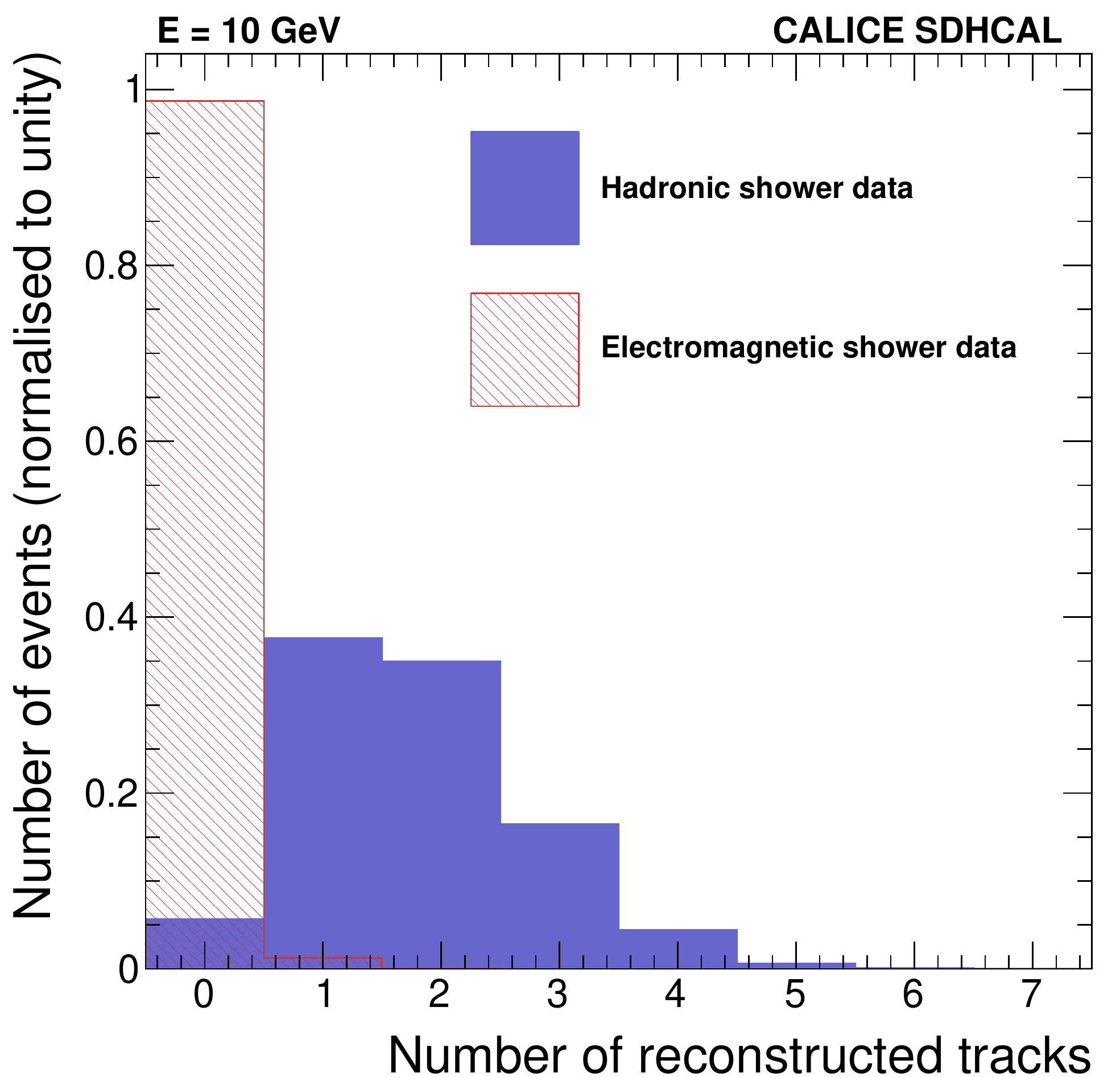}
\includegraphics[width=.32\textwidth]{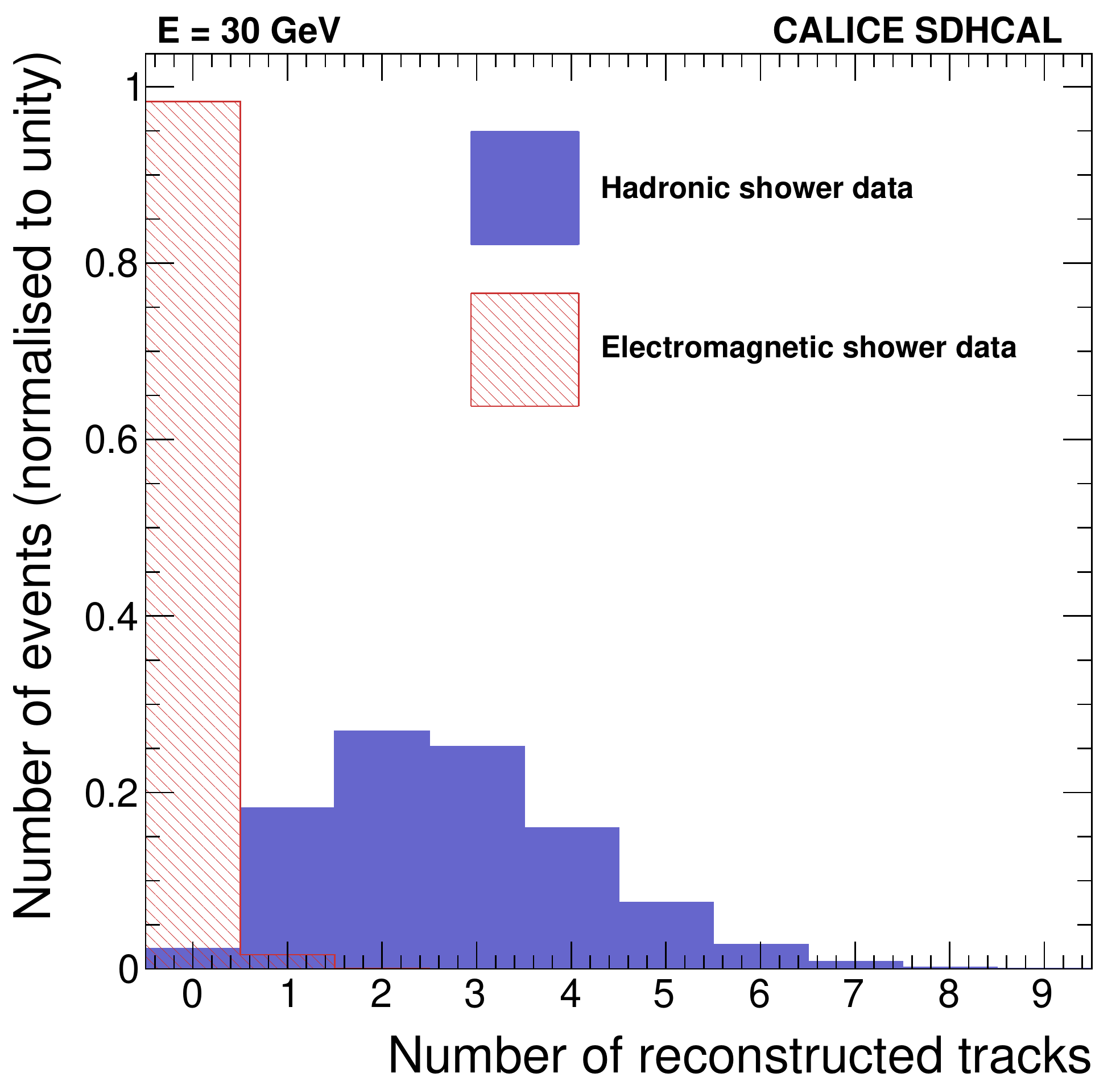}
\includegraphics[width=.32\textwidth]{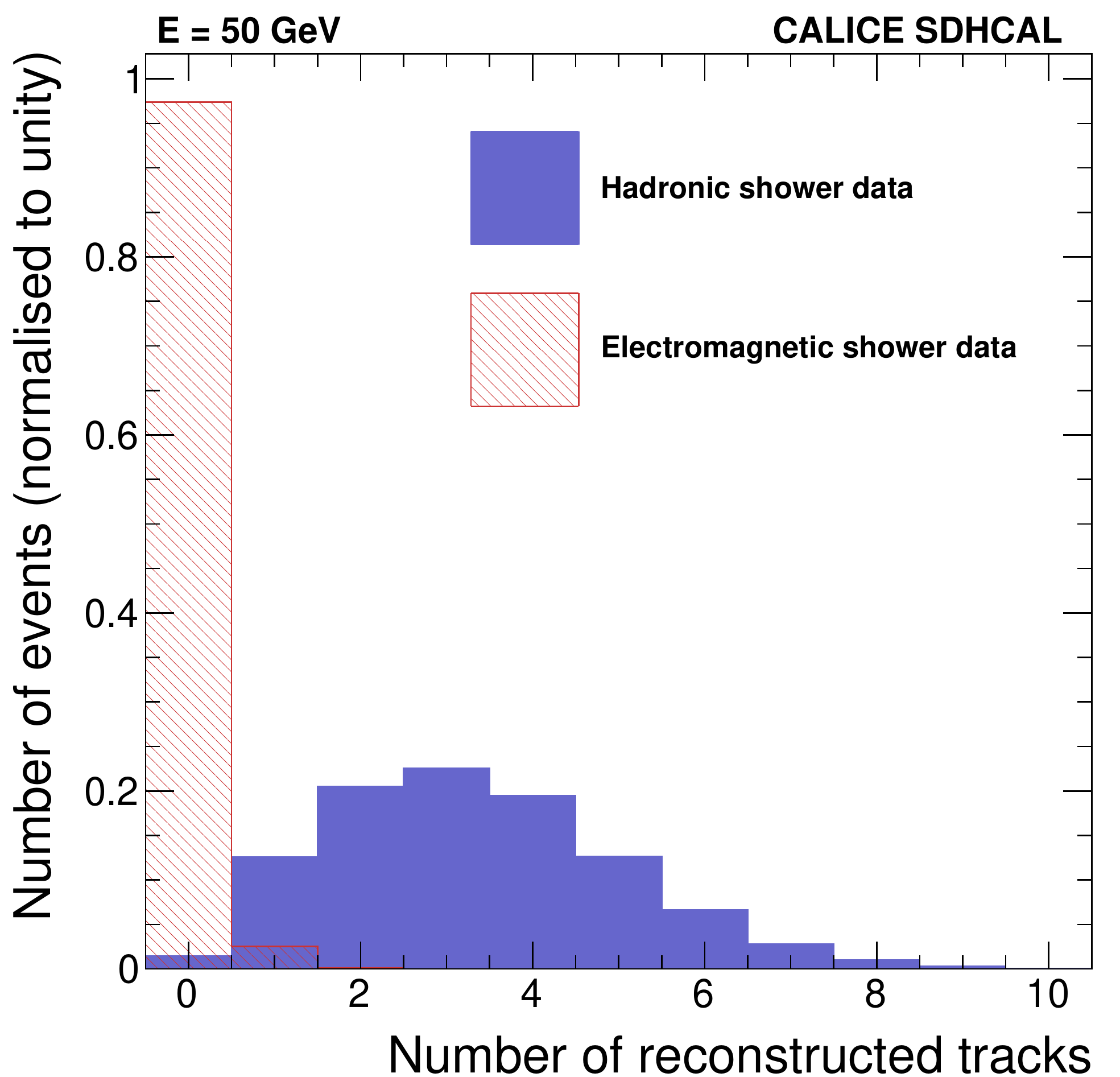}
\caption{Number of reconstructed tracks in showers produced by electron (hatched histogram) and pions (filled histogram) at 10, 30 and 50 GeV.}
\label{fig:elecTrack}
\end{figure}

The same HT track reconstruction is applied to  events collected  with  electron beams in the 10 to 50 GeV  energy range using a special filter during the 2012 SDHCAL beam test as explained in ref.~\cite{sdhcal-paper}.   The absence, as expected in electromagnetic shower,  of such tracks in the case of electrons,  compared with pions as shown in figure~\ref{fig:elecTrack} shows the low probability of the HT method to introduce fake tracks.

It is worth noting here that  this  difference in number of track segments between electromagnetic and hadronic showers can also be used to discriminate the two species for particle identification purposes\footnote{A future paper will be dedicated to  electron and hadron separation in the SDHCAL.}. Requiring at least one track segment rejects indeed 99\% (97\%) of electromagnetic showers while keeping more than 95\% (99\%) of the hadronic showers at 10 (50) GeV respectively.

\section{ Tracks in hadronic shower models}\label{models}
The track segments produced in showers collected in the SDHCAL prototype can be used as a tool to compare different hadronic shower models used in the simulation. The number of tracks and their characteristics are related to those of charged particles (pions, kaons and  protons) produced in the hadronic shower with an energy sufficient to cross a few absorber layers.  In absence of high-granularity calorimeters such variables could not be easily used to tune the phenomenological models.  Studying these segments in SDHCAL thus constitutes an unbiased tool to compare  among the different models on the one hand and with the data on the other hand. 
 Events with different energies produced by pion interactions in the SDHCAL prototype were simulated using  a few hadronic shower models within the  version 9.6p01 of the GEANT4 framework~\cite{g4-status}.  
A  digitizer\cite{digitizer}  transforms the energy deposited by the particles crossing the active volumes into charges and induced signal in the neighbouring pads.  In the case of a single charged particle crossing one pad of an active layer,  the digitizer's parameters are  tuned to reproduce the efficiency and the pad multiplicity observed with beam muons. Cosmic muons are also used for this purpose to simulate correctly the track segments produced in hadronic showers at large angles with respect to the incoming hadron. 
The parameters are also optimized to reproduce  the response of the GRPC to the passage of several charged particles in one pad by taking into account the charge screening effect. To do so, only the electron data are used in order to avoid biases when comparing the data with the simulation of the hadronic shower models.  The same set of parameters is then used to simulate pion showers. 

\begin{figure}[!ht]
\includegraphics[width=.32\textwidth]{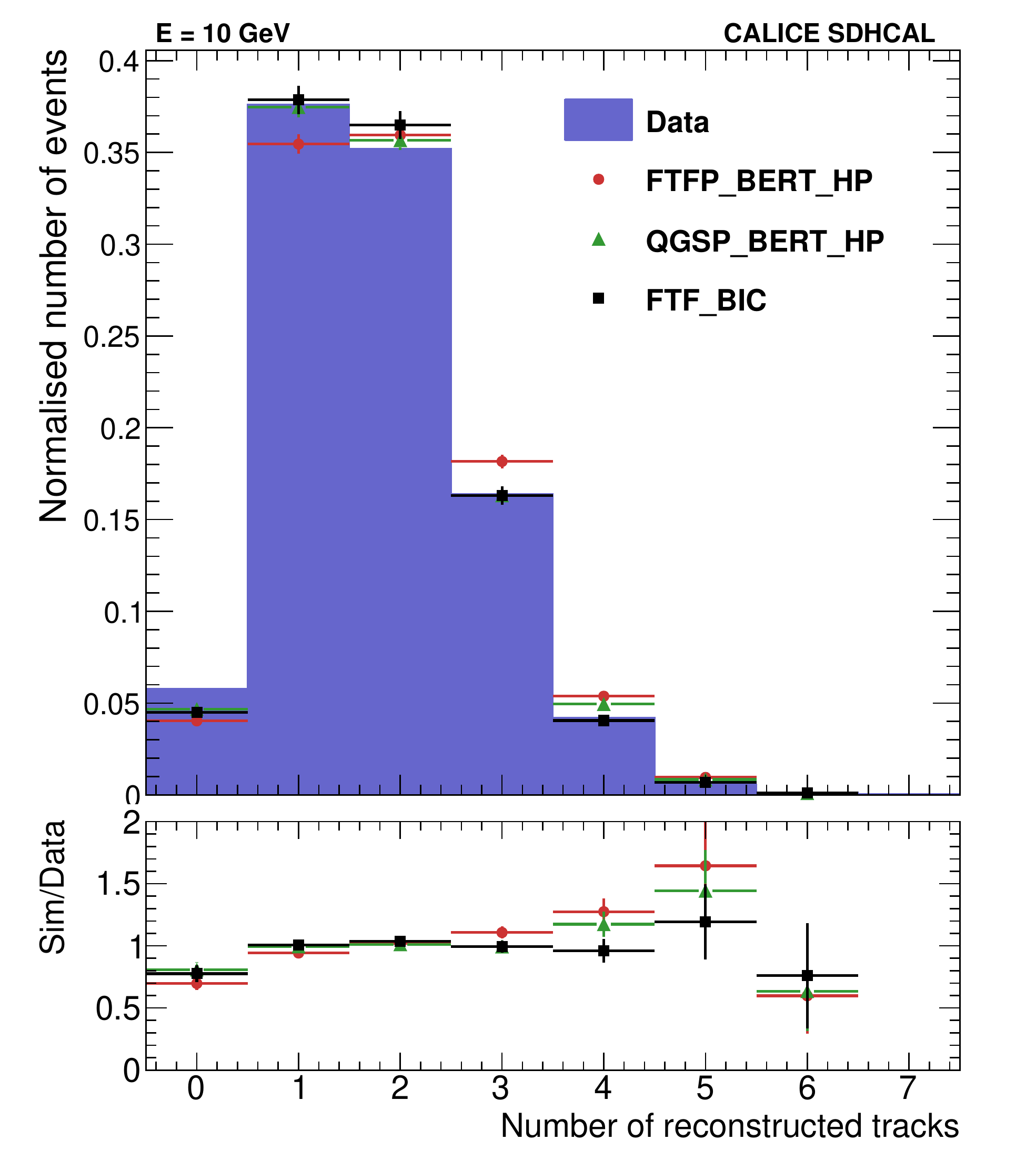}
\includegraphics[width=.32\textwidth]{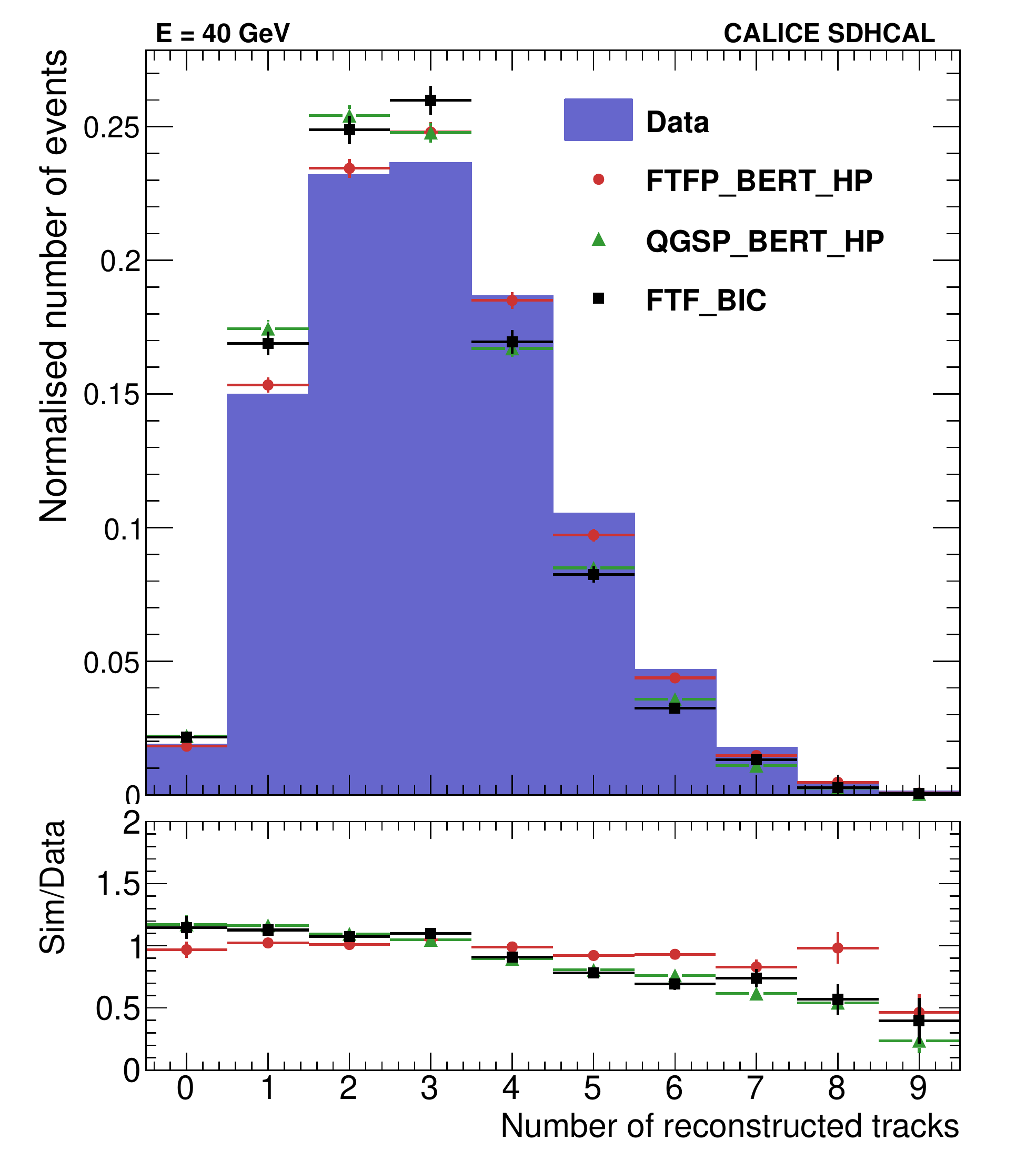}
\includegraphics[width=.32\textwidth]{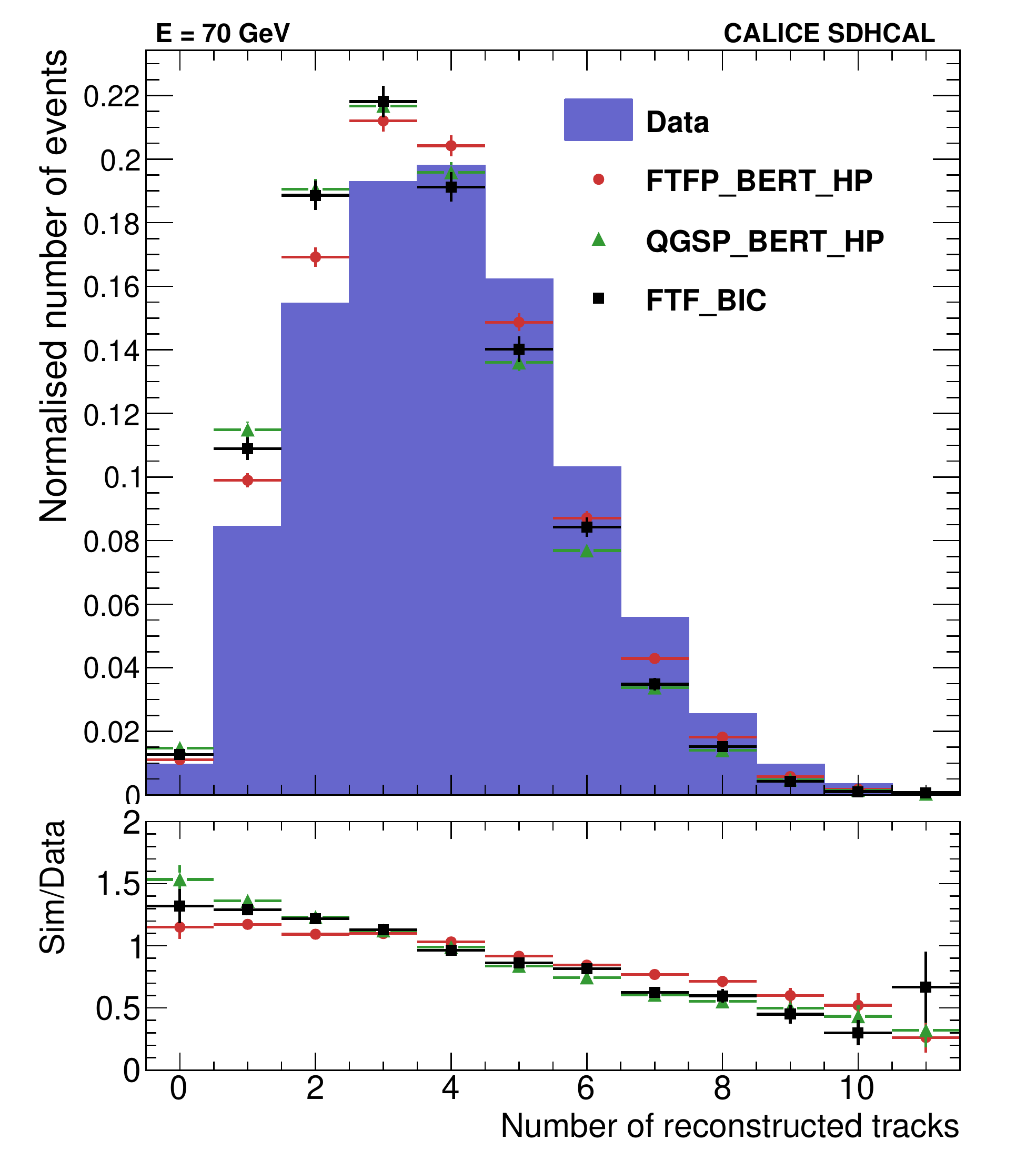}
\caption{Top: distribution of the number of the reconstructed track segments in hadronic shower for simulation and for data at 10, 40 and 70 GeV. Bottom: ratio of the same distribution between the simulation and data.  Only statistical uncertainties are included.}
\label{fig:trackmulti}
\end{figure}

\begin{figure}[!ht]
\includegraphics[width=.32\textwidth]{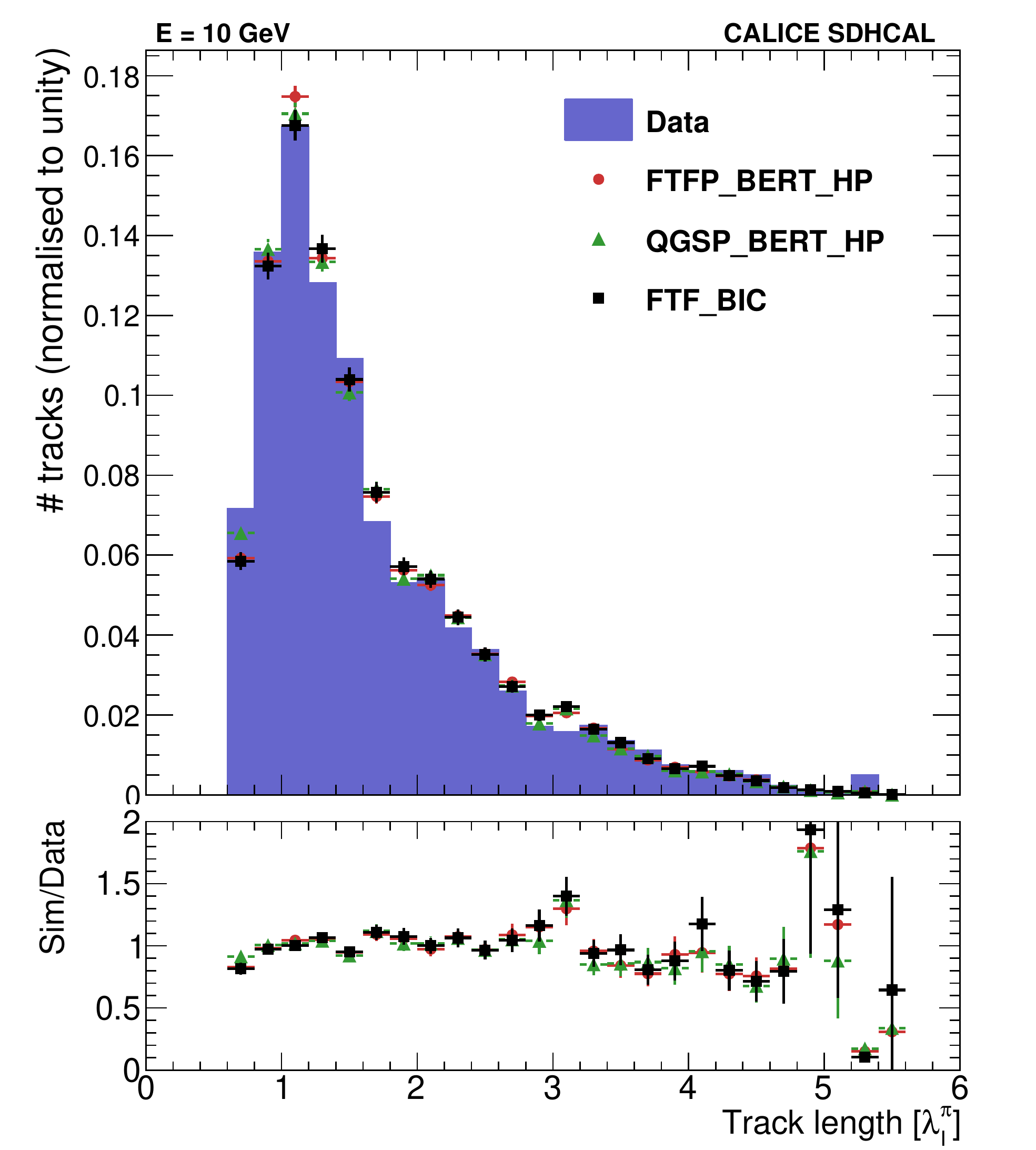}
\includegraphics[width=.32\textwidth]{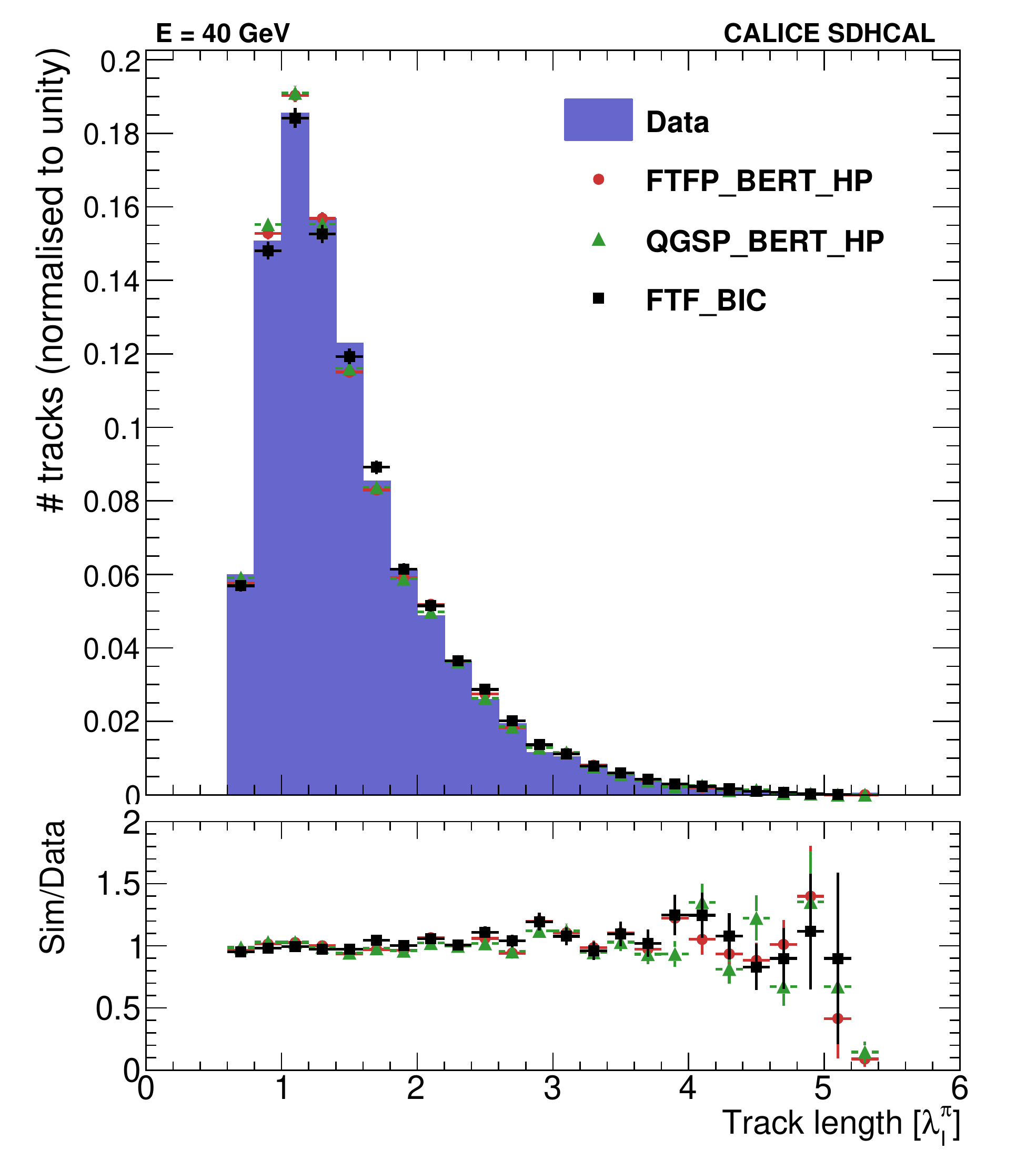}
\includegraphics[width=.32\textwidth]{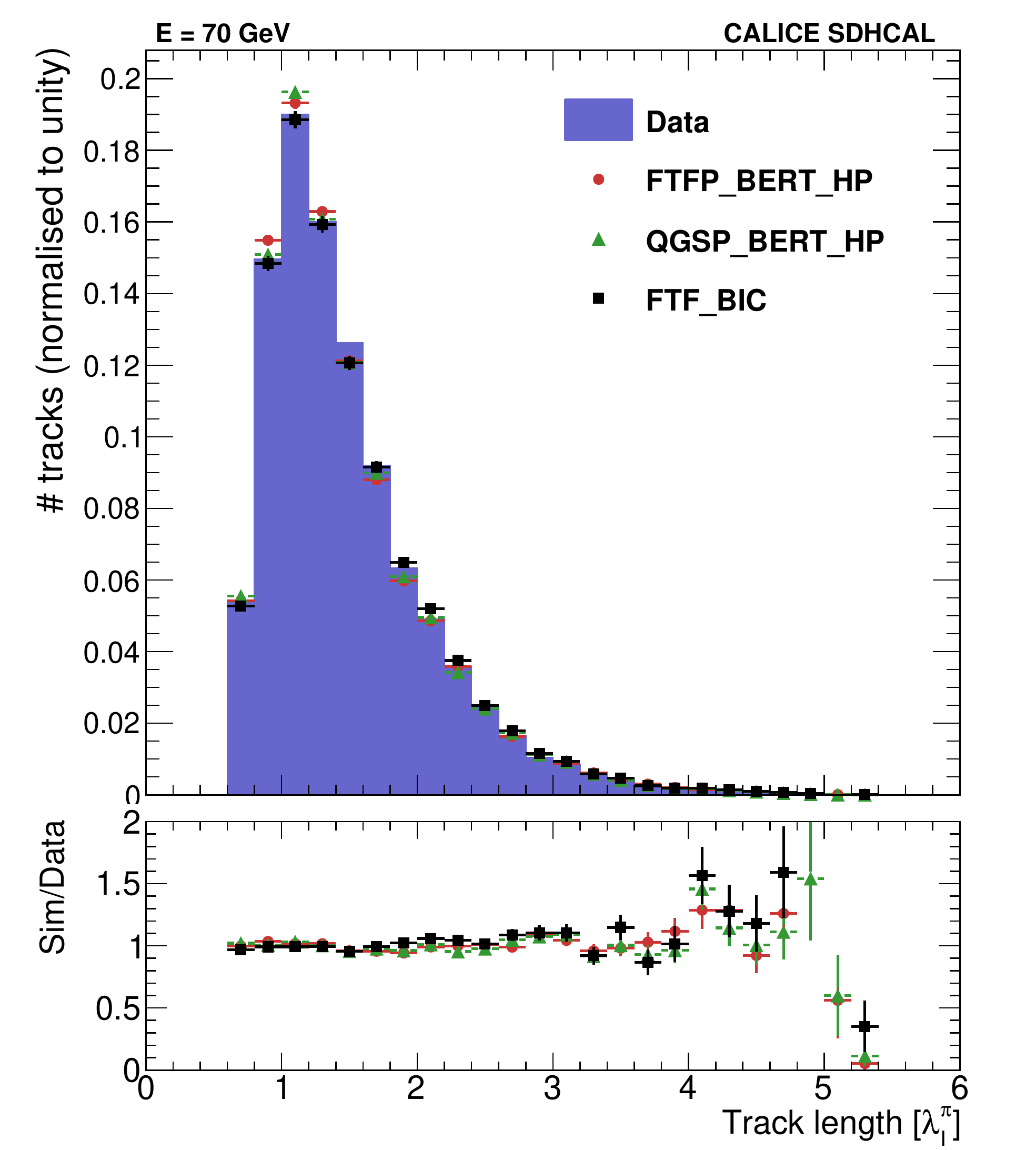}
\caption{Top: distribution of the track segment length in hadronic showers for simulation and for data at 10, 40 and 70 GeV. The length is given in interaction length units. Bottom: ratio of  the same distribution  between the simulation and data. Only statistical uncertainties are included.}
\label{fig:tracklength}
\end{figure}

\begin{figure}[!ht]
\includegraphics[width=.32\textwidth]{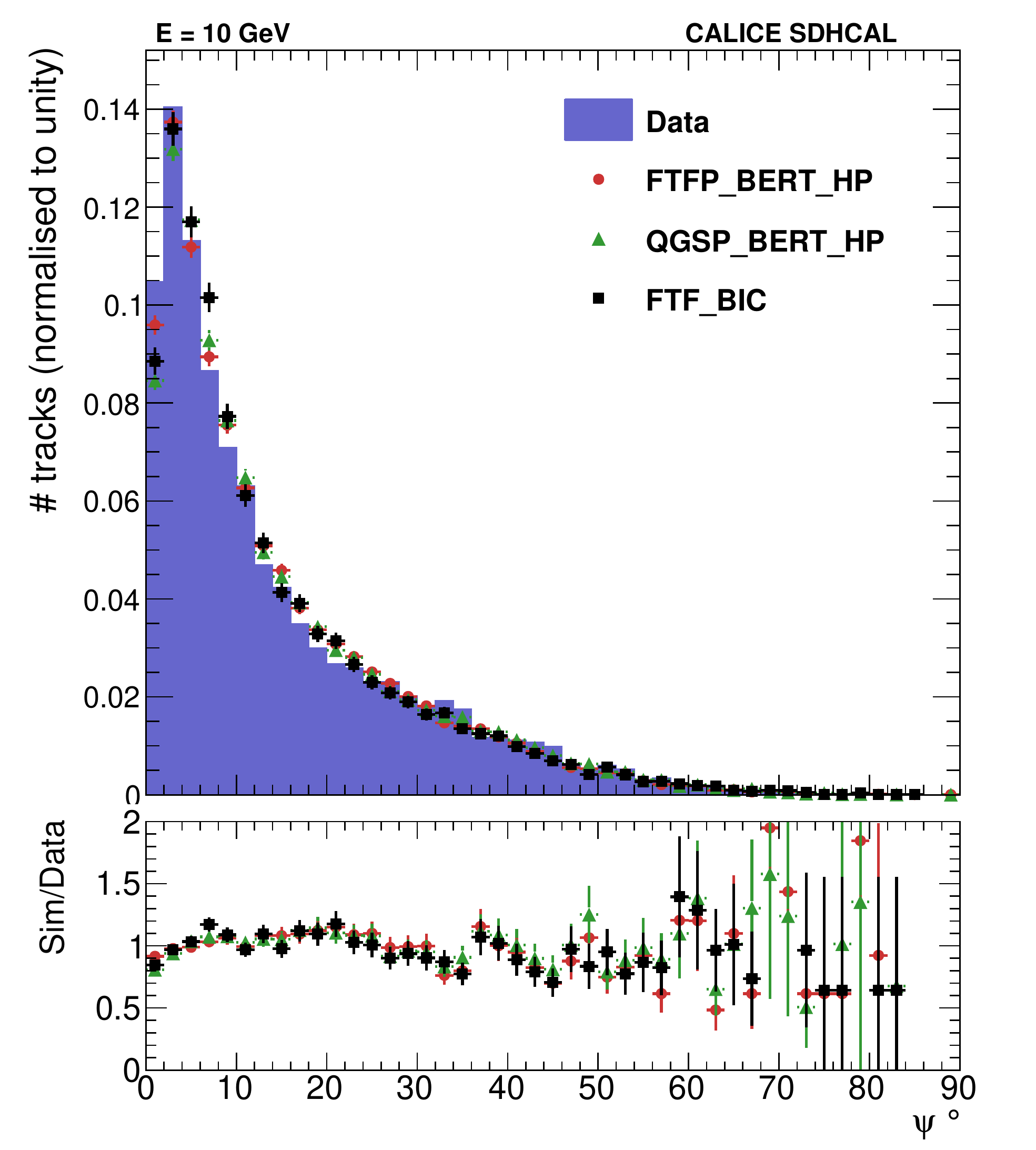}
\includegraphics[width=.32\textwidth]{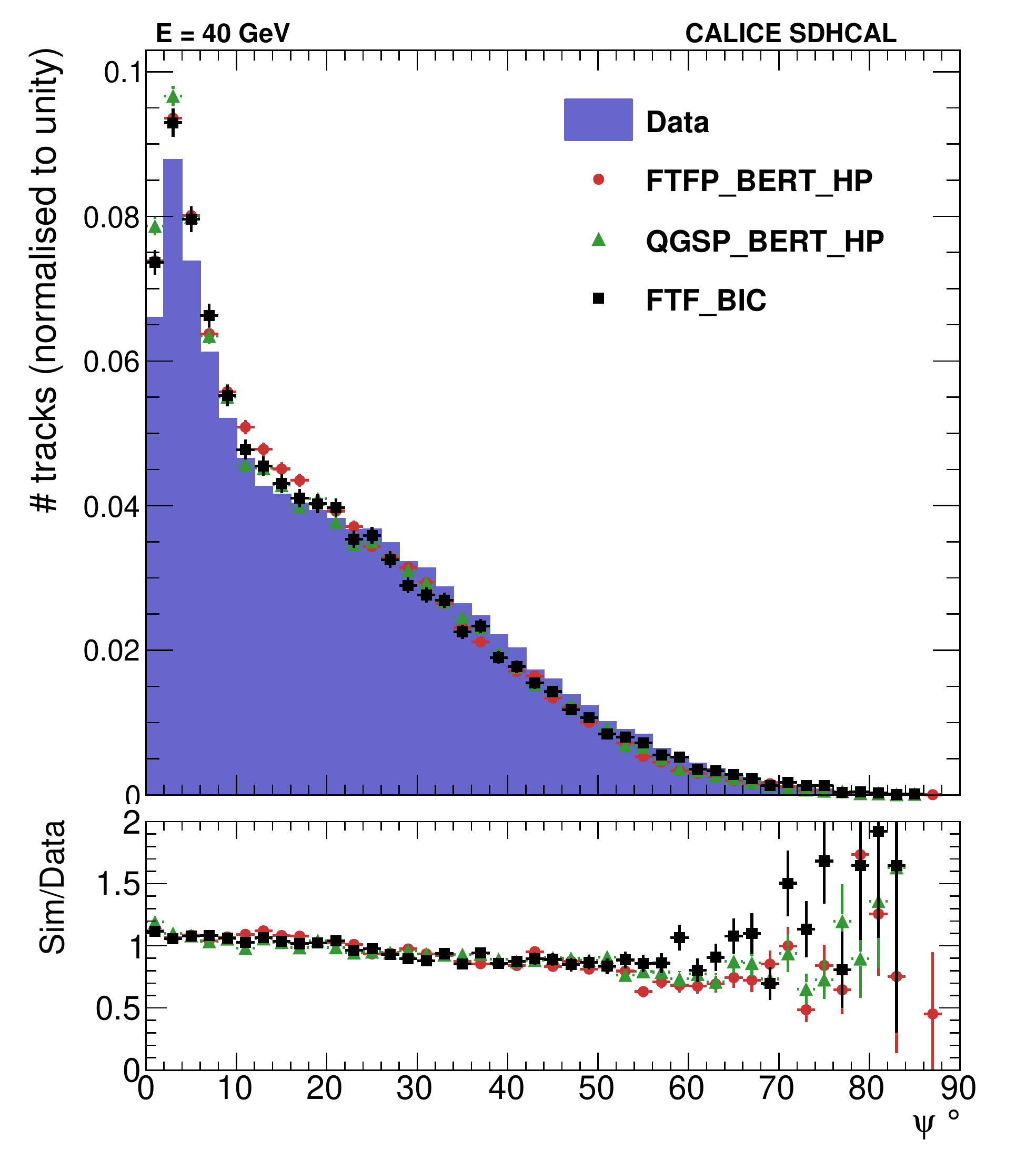}
\includegraphics[width=.32\textwidth]{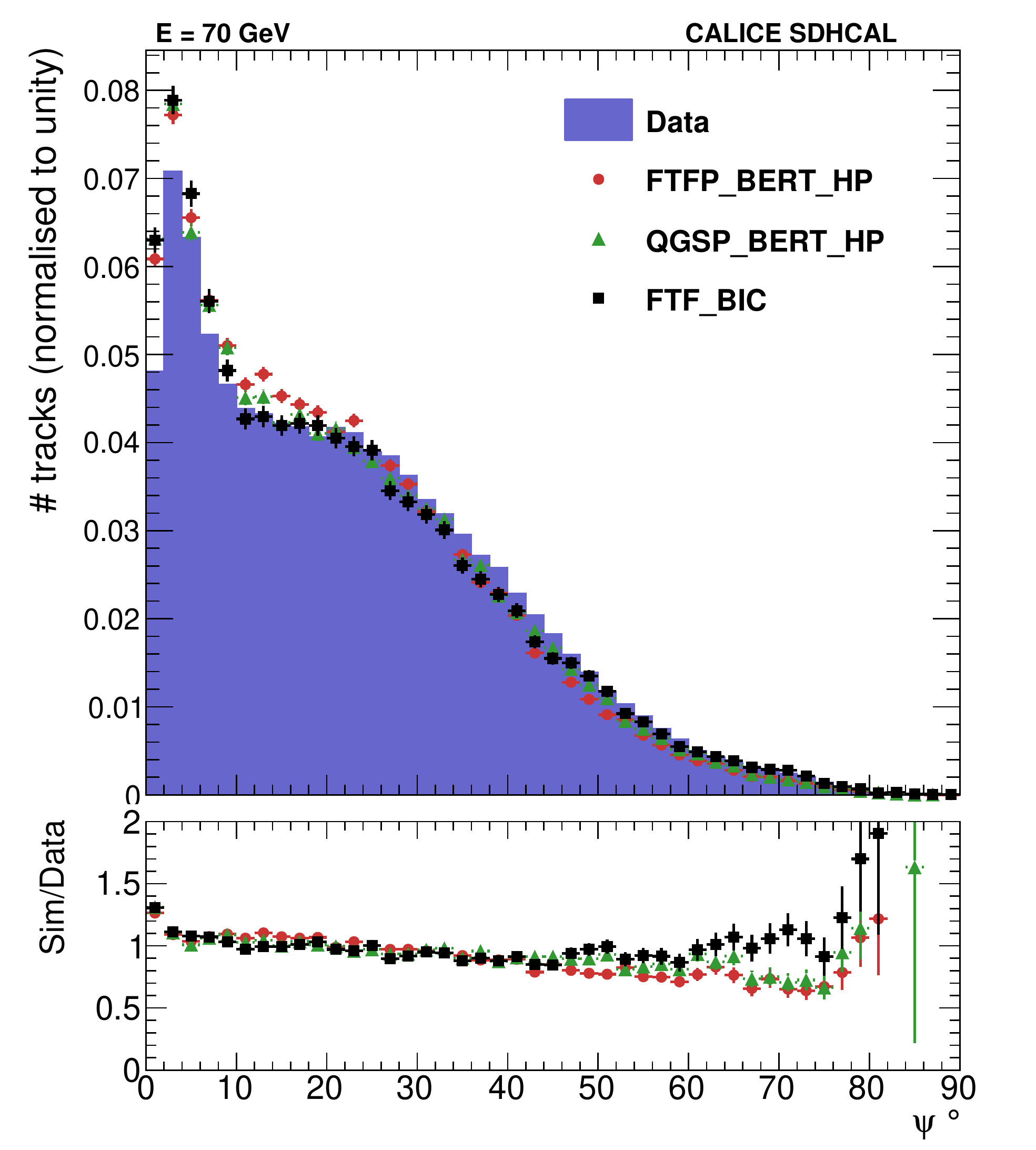}
\caption{Top: distribution of  the angle of the track segments in hadronic showers with respect to the incoming hadron for simulation and for data at 10, 40 and 70 GeV. Bottom: ratio of  the same distribution between the simulation and data. Only statistical uncertainties are included.}
\label{fig:trackangle}
\end{figure}

\begin{figure}[!ht]
\includegraphics[width=.32\textwidth]{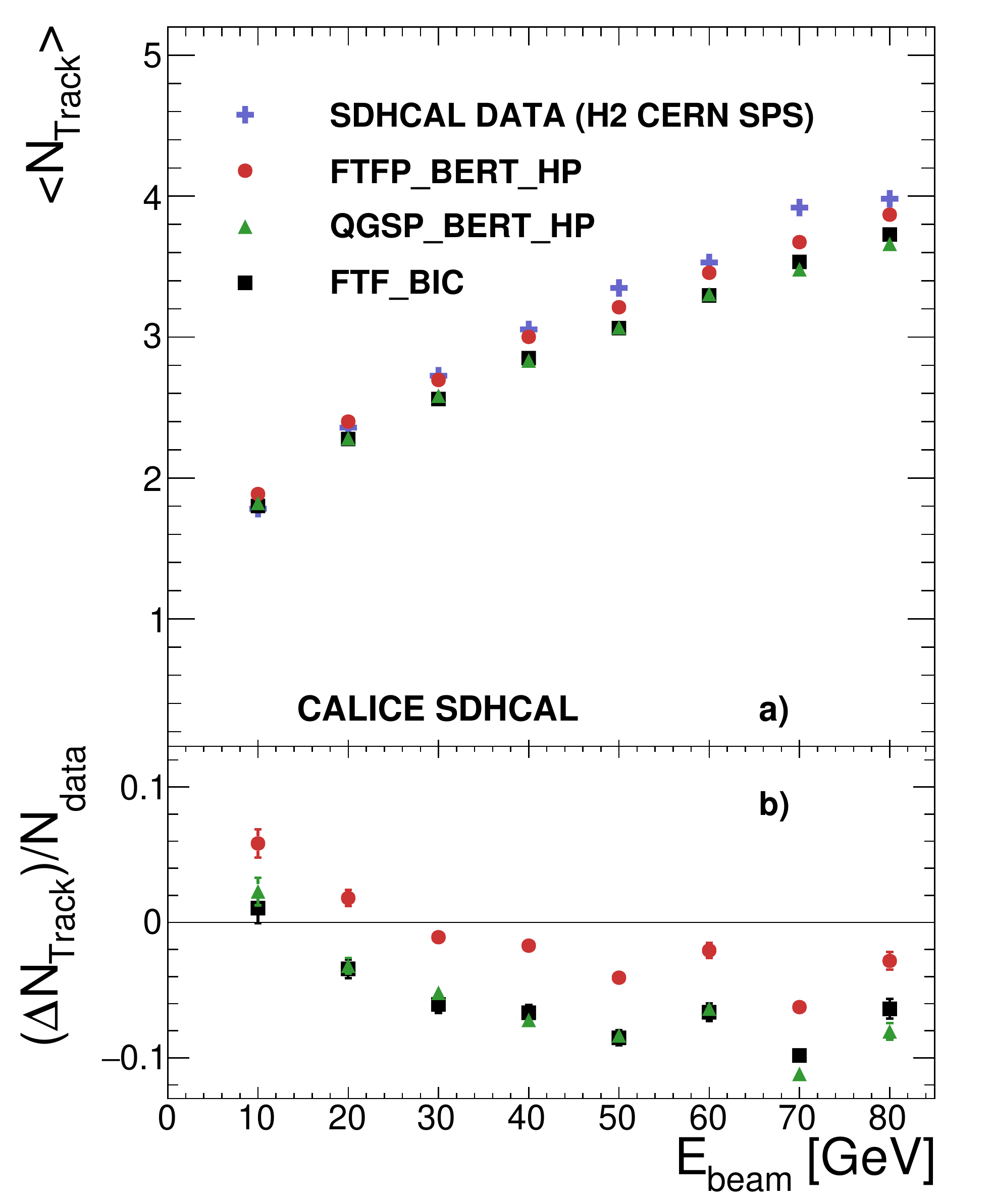}
\includegraphics[width=.32\textwidth]{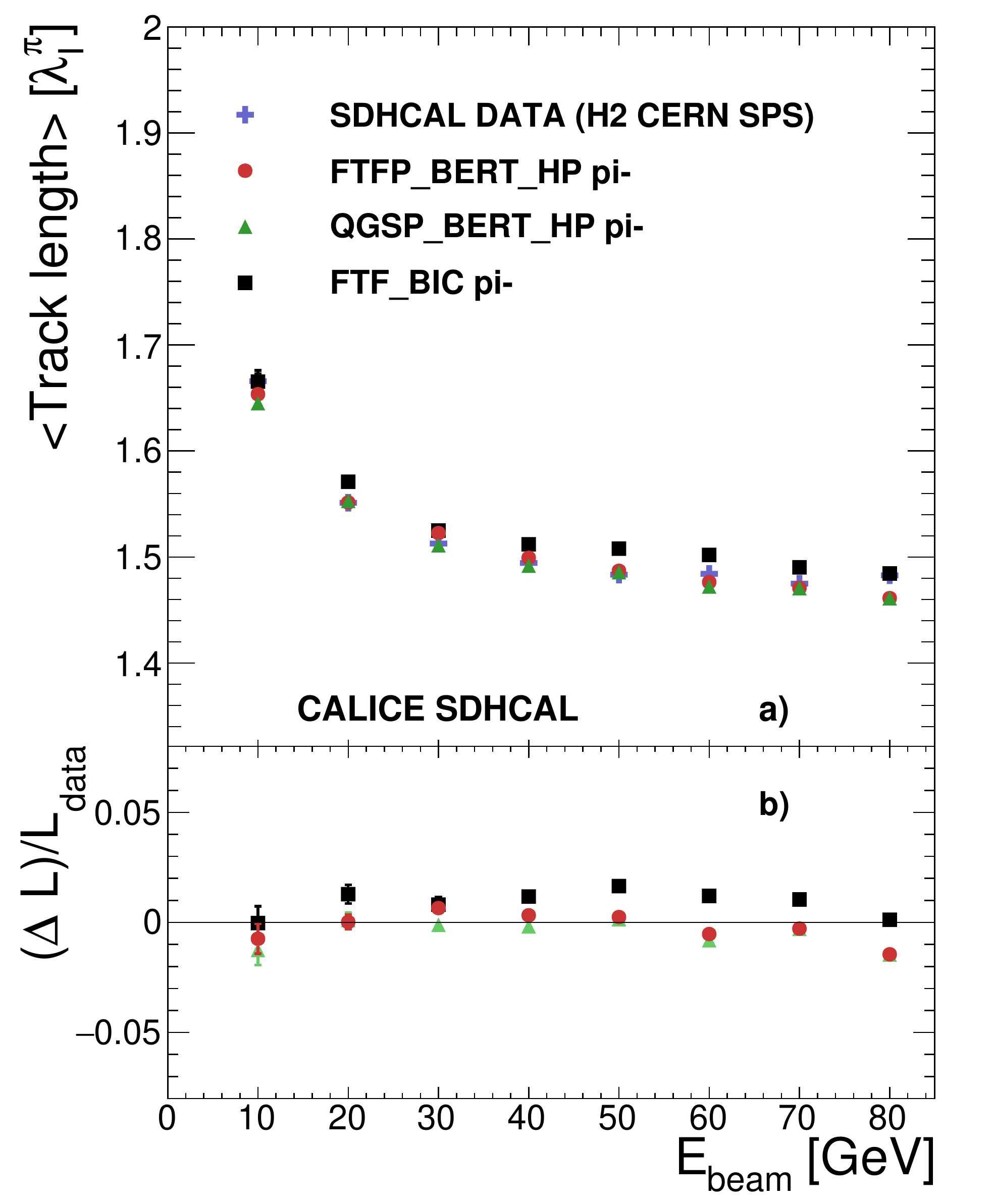}
\includegraphics[width=.32\textwidth]{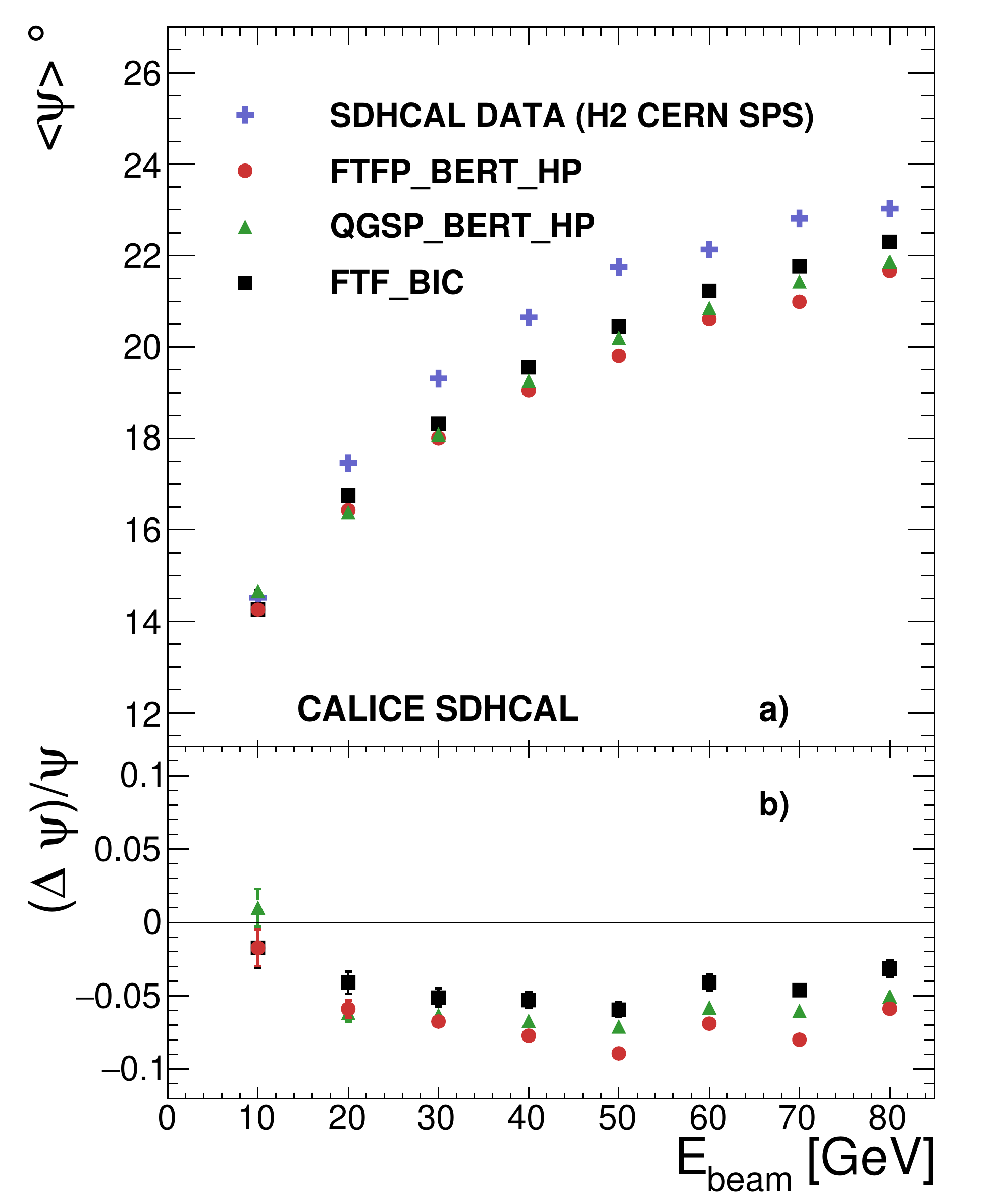}
\caption{Left: Mean number of reconstructed track segments in hadronic showers as a function of the beam energy (a) and the relative difference between simulation and data (b). Middle: Mean track segment length as a function of the beam energy (a) and the relative difference between simulation and data (b). Right: Mean angle of the track segments  as a function of the beam energy (a) and the relative difference between simulation and data (b). Only statistical uncertainties are included.} 
\label{fig:trackMulti-Length-Angle}
\end{figure}

Three phenomenological models \texttt{FTF\_BERT\_HP},  
 \texttt{QGSP\_BERT\_HP}  and  \texttt{FTF\_BIC} are studied. The tracks  obtained using the HT in simulated events with these three  models are  compared to each other and to data for different energies.  The distributions of the total number of reconstructed tracks within 10, 40 and 70 GeV hadronic showers are shown in figure~\ref{fig:trackmulti}. The track length could be an interesting variable to compare simulation models with data. It is defined as the distance between the most upstream and downstream  of the clusters belonging to  a given HT selected  track segment.  Figure~\ref{fig:tracklength} shows the track length distribution of the data and the simulated events for the three energies. 
Another feature that may help to discriminate  between the different hadronic models is the angular distribution of the track segments with respect to the incoming hadron. To determine the angle $\psi$ of the track segments with the incoming hadron, the direction of the latter was determined by using a linear fit of the  barycentre coordinates of the clusters in the first ten layers.  
 The angular distribution of the track segments found with the HT method for the same three energies is shown in figure~\ref{fig:trackangle}.
Finally, in  figure~\ref{fig:trackMulti-Length-Angle} are shown the average number of track segments (left), their average  length (middle) and  their average  angle (right) with respect to the incoming pion as a function of the beam energy as obtained in simulated events using each of the three models and in data events.  
The three models seem to reproduce fairly well the number of reconstructed track segments observed in data with \texttt{FTF\_BERT\_HP} providing a better description at high energy but worse at 10 GeV.  In the same way, \texttt{FTF\_BIC} features slightly longer track segments  with respect to the other models when compared with the data at high energy. All three models fail to describe adequately the angular distribution of the track segments.  These results are in agreement with those found in the study of track segments in the CALICE AHCAL prototype~\cite{AHCALSegment} although here the model  \texttt{FTF\_BERT\_HP} seems to describe slightly better the number of segments than the \texttt{QGSP\_BERT\_HP} while in ref.~\cite{AHCALSegment} the opposite is observed. This difference could be explained by the fact that the version 9.4p02  is used in in ref.~\cite{AHCALSegment}  while here  we use a more recent version (9.6p01) of the \texttt{FTF\_BERT\_HP}.  The use of the version \texttt{QGSP\_BERT\_HP} (respectively (\texttt{FTF\_BERT\_HP})  here and the   \texttt{QGSP\_BERT} (respectively \texttt{FTF\_BERT})  in ref.~\cite{AHCALSegment} should not in principle  impact the comparison since the difference between  the two versions is essentially related to the treatment of neutrons which is not relevant in this study.

In the present comparison only statistical uncertainties are included.  Some sources of the systematic uncertainties such as the ones related to the minimum number of clusters required to apply the HT selection and the $(\theta, \rho)$  histogram binning are studied  and found to be negligible.   Although more detailed systematics study is needed, the low noise of the SDHCAL~\cite{sdhcal-paper}\footnote{ Only one noise hit is expected in a physics event in the SDHCAL prototype. This is to be compared with an average number of 200 (1500) hits  in a 10 (80)~GeV pion shower respectively.}, the fact that the  track efficiency per layer in data is well reproduced in the simulation~\cite{digitizer}  suggest that the contribution of the other systematic uncertainties will not modify the conclusion of this study.  The impact of the difference in the track multiplicity which varies slightly from one layer to other in data while it is  almost constant in the simulation is  absorbed since clusters rather than the hits are used here to build the track segments in a low dense environment.



\section{Conclusion} 
The Hough Transform is a simple and powerful method for finding track segments  within a noisy environment. A new technique to use this method in hadronic showers is developed and successfully applied to events collected  during the exposure of the CALICE SDHCAL to hadron beams. 
The advantages of using track segments obtained with this technique to calibrate the hadronic calorimeter in situ are shown.  A slight improvement on the energy reconstruction is also obtained by giving the same weight to the hits belonging to track segments irrespective of  their threshold. 

The same technique are also applied to simulated hadronic showers. Comparison with data helps to better characterize the different hadronic shower models used in the simulation. \texttt{FTF\_BERT\_HP} model seems to be  slightly closer to the SDHCAL  data than  the  \texttt{QGSP\_BERT\_HP} and  \texttt{FTF\_BIC} ones.

The extension of the method  to hadronic showers in the presence of magnetic field should complete this work and allows the technique to be used in highly granular calorimeters in future experiments. 

\section{Acknowledgements}
We would like to thank the CERN-SPS staff for their availability and precious help during the beam test period.   We would like to acknowledge the important support provided by the  F.R.S.-FNRS, FWO (Belgium), CNRS and ANR (France), SEIDI and CPAN (Spain). This work was also supported by the Bundesministerium f\"{u}r Bildung und  Forschung (BMBF), Germany; by the Deutsche Forschungsgemeinschaft  (DFG), Germany; by the Helmholtz-Gemeinschaft (HGF), Germany; by  the Alexander von Humboldt Stiftung (AvH), Germany;  by the Korea-EU cooperation programme of National Research Foundation of Korea, Grant Agreement 2014K1A3A7A03075053
;by the National Research Foundation of Korea.

\end{document}